\documentclass[reprint,amsmath,amssymb,showkeys,showpacs,floatfix]{revtex4-2}
\usepackage[utf8]{inputenc}
\usepackage{graphicx}
\usepackage{color}
\usepackage{booktabs}
\usepackage{array}[=2016-10-06]
\usepackage{amsfonts} 
\usepackage{ragged2e}
\usepackage{tabularx}
\usepackage[colorlinks=true,linkcolor=blue,citecolor=blue,urlcolor=blue]{hyperref}

\newcommand{\ket}[1]{\mathinner{|{#1}\rangle}}
\newcommand{\bra}[1]{\mathinner{\langle{#1}|}}

\begin{document}
\title{Spin-Polarized Transport Tomography of a Quantum-Dot Spin Qubit via Ferromagnetic Leads}
\author{M. B. Samb\'u} 
\affiliation{QPequi Group, Institute of Physics, Federal University of Goi\'as, Goi\^ania, Goi\'as, 74690-900, Brazil}
\affiliation{Instituto de F\'isica, Universidade Federal de Uberl\^andia, 38400-902 Uberl\^andia, MG, Brazil}
\author{L. Sanz}
\email{lsanz@ufu.br}
\author{F. M. Souza}
\email{fmsouza@ufu.br}
\affiliation{Instituto de F\'isica, Universidade Federal de Uberl\^andia, 38400-902 Uberl\^andia, MG, Brazil}

\begin{abstract}
We propose a theoretical protocol for reconstructing the density matrix of a single-electron spin qubit using spin-polarized transport. The system consists of a quantum dot coupled to ferromagnetic reservoirs and subject to a magnetic field lying in the $xy$ plane of the Bloch sphere. Spin-dependent tunneling events measured along the $x\pm$, $y\pm$, and $z\pm$ quantization axes give rise to probability distributions that encode the quantum state of the qubit.
The open-system dynamics are described using a Lindblad master equation, which captures the time evolution of the spin under continuous coupling to the reservoirs. By counting tunneling events for four different magnetic alignments, we formulate a scheme for reconstructing the full density matrix of the qubit. The resulting simulation data are analyzed using Gaussian Process Regression (GPR) as a supervised machine-learning post-processing tool to reconstruct smooth time-dependent density-matrix elements from the measured probability distributions. The proposed model enables complete access to the open-system density matrix, including both population probabilities and relative phase information. Successful state reconstruction demonstrates the validity and robustness of the approach, highlighting its applicability to experimentally accessible spin-transport platforms.
\end{abstract}
\maketitle
\section{Introduction}
\label{sec:intro}
Since the seminal work of Loss and DiVincenzo on quantum computation with single spins in quantum dots~\cite{loss1998}, a wealth of experimental and theoretical studies on quantum information processing with quantum bits (qubits) based on electron spin have emerged in the literature~\cite{zhang2018}, mainly motivated by the long spin decoherence times for electrons in semiconductors and the ability to measure and control quantum-dot spins~\cite{yoneda2018}. A few examples include the implementation of CNOT gates in semiconductor quantum dots using electron spin qubits~\cite{zajac2018}, the $\sqrt{SWAP}$ two-qubit exchange gate between electron spin qubits~\cite{he2019}, and both CZ and CNOT gates in double quantum dot structures~\cite{russ2018}. Recently, a valley-orbit hybridization mechanism was shown to protect electronic qubits in quantum dots from the effects of charge noise~\cite{mi2018}. Recent progress and perspectives on semiconductor spin qubits are reviewed in Ref.~\cite{Burkard2023}.

The reconstruction of qubit states is of fundamental importance for quantum technologies, as their characterization enables the study of properties such as the probabilities of possible measurable outcomes, coherences, and entanglement~\cite{cramer2010}. The field of quantum state tomography emerged within the context of quantum optics, with the development of methods to reconstruct photonic states~\cite{vogel1989, smithey1993, ariano1995, schiller1996}, with modern approaches extending these ideas to complex quantum systems and scalable reconstruction schemes~\cite{Carrasquilla2019}.

In the field of quantum information carried by particles in quantum dots, various efforts in tomography have been employed to characterize the state of the system. Existing tomography protocols rely on different physical mechanisms and therefore exhibit distinct advantages and limitations. For instance, time-resolved Kerr rotation spectroscopy facilitates non-destructive observation of spin precession in a quantum dot~\cite{mikkelsen2007}. The tomographic Kerr rotation technique has been refined to observe the spin states of optically injected electrons in a semiconductor quantum well~\cite{kosaka2009}. Tomographic methods have been extended to two- and three-partite electronic systems~\cite{shulman2012,medford2013}. Additionally, leveraging shot-noise, Jullien \emph{et al.} reconstructed the wavefunction of single electrons in ballistic conductors by repeatedly applying Lorentzian voltage pulses to inject on-demand electrons into a conductor~\cite{jullien2014}. Also, Bisognin \emph{et al.}~\cite{bisognin2019} demonstrated fermionic quantum state reconstruction from electrical currents, thereby opening new possibilities for quantum transport-based tomography. More recently, Cogan et al. reported the complete state tomography of \textit{an anchored} quantum-dot single electron spin  qubit using an all-optical protocol based on resonant excitation and polarization-resolved photon detection~\cite{Cogan2020}. Still, optical techniques provide high-fidelity state reconstruction but require optical access and resonant excitation of the qubit. Conversely, transport-based approaches are naturally compatible with electrically controlled spintronic architectures and can exploit the stochastic tunneling events that are already measured during device operation. A comparison between optical and transport approaches is presented in Table~\ref{tab:comparison}. 
\begin{table*}[ht]
	\caption{Comparison between representative state-characterization protocols for semiconductor electronic systems and the transport-based tomography scheme proposed in this work.}
	\begin{ruledtabular}
		\begin{tabular}{ 
				>{\raggedright\arraybackslash}p{0.22\linewidth} 
				>{\raggedright\arraybackslash}p{0.22\linewidth} 
				>{\raggedright\arraybackslash}p{0.15\linewidth} 
				>{\raggedright\arraybackslash}p{0.16\linewidth} 
				>{\raggedright\arraybackslash}p{0.18\linewidth} 
			}
			Method &
			Measurement principle &
			Platform &
			Target state characterization &
			Main features \\
			\colrule
			
			Time-resolved Kerr rotation~\cite{mikkelsen2007,kosaka2009} &
			Optical Kerr rotation &
			Semiconductor quantum wells / quantum dots &
			Electron spin dynamics and tomography &
			Non-destructive optical probing; requires optical access \\
			
			Tomographic protocols for multipartite spin systems~\cite{shulman2012,medford2013} &
			Spin measurements after controlled operations &
			Coupled quantum-dot spin qubits &
			Two- and three-qubit quantum states &
			Requires coherent gate operations and multiple measurement bases \\
			
			Single-electron wavefunction reconstruction~\cite{jullien2014} &
			Shot-noise interferometry &
			Ballistic conductors &
			Single-electron wavefunction &
			Transport-based characterization of propagating electrons \\
			
			Fermionic quantum-state reconstruction~\cite{bisognin2019} &
			Electrical current correlations &
			Mesoscopic conductors &
			Fermionic quantum states &
			Electrical reconstruction from current measurements \\
			
			All-optical spin tomography~\cite{Cogan2020} &
			Resonant optical excitation and polarization-resolved detection &
			Quantum-dot spin qubit &
			Complete single-spin density matrix &
			High-fidelity optical protocol requiring resonant optical control \\
			
			\textbf{Present work: Spintronic tomography} &
			Spin-dependent tunneling statistics through ferromagnetic leads &
			Open quantum-dot single-electron spin qubit &
			Complete reduced density matrix &
			Transport-based protocol formulated within a Lindblad framework and naturally compatible with electrically controlled spintronic architectures \\
		\end{tabular}
	\end{ruledtabular}
	\label{tab:comparison}
\end{table*}

The protocol proposed in this work follows this latter philosophy: a transport-based tomography protocol for an open quantum-dot single-electron spin qubit, in which the quantum state is reconstructed from spin-dependent tunneling statistics through ferromagnetic leads with different magnetization directions. It therefore provides a conceptually different route to quantum state reconstruction that is naturally suited to spintronic architectures based on electrically detected transport. The system under investigation consists of a spintronic device based on a quantum dot coupled to ferromagnetic leads~\cite{souza2007}. We focus on spin-dependent tunneling events measured along the $x\pm$, $y\pm$, and $z\pm$ quantization directions, which generate probability distributions carrying complete information about the qubit state. The method is formulated within the Lindblad description of open quantum systems and exploits repeated spin-polarized measurements to recover both the populations and coherences of the reduced density matrix. 

By systematically probing different magnetic configurations and repeating identical initialization and detection sequences, we show that the full density matrix of a single qubit can be reconstructed, including both populations and coherences. The statistics employed for the reconstruction are obtained from stochastic simulations of the open-system dynamics, closely mimicking realistic experimental measurement outcomes. Supervised machine-learning techniques are then employed as a post-processing tool to reconstruct smooth time-dependent density-matrix elements from the simulated counting statistics. Machine-learning-assisted quantum state reconstruction has recently attracted increasing attention as a powerful tool for analyzing complex measurement data~\cite{Innan2024}.

Among the supervised machine-learning methods available for 
post-processing tomographic data, Gaussian Process Regression 
(GPR)~\cite{rasmussen2006} is particularly well suited for the 
present application for several reasons. First, unlike polynomial 
regression or neural networks, GPR performs well with sparse and 
irregular training data --- precisely the situation arising from 
stochastic detection events, which occur only at a finite and 
randomly distributed subset of time points. Second, GPR provides 
native quantified uncertainty bands ($\pm 1\sigma$) at every 
predicted time point, reflecting the statistical quality of the 
detection data --- a feature not available with polynomial 
regression, random forests, or standard neural networks without 
additional ensemble techniques. Third, the GPR kernel can be chosen 
to encode prior physical knowledge about the signal structure: the 
squared-exponential (RBF) kernel captures the smooth oscillatory 
behavior characteristic of Rabi dynamics, while the WhiteKernel 
explicitly models the stochastic noise of the detection process. 
Finally, the kernel hyperparameters are optimized automatically via 
maximization of the log marginal likelihood, removing the need for 
ad hoc choices such as the polynomial degree.

The paper is organized as follows. In Sec.~\ref{sec:theory}, we present the theoretical framework of the proposed model. Section~\ref{sec:simulations} is devoted to the simulation of spin-polarized transport experiments for single-qubit state tomography. In Sec.~\ref{sec:randomdata}, we present and discuss the results of the density-matrix reconstruction obtained from the simulated tunneling statistics. Finally, our main results and conclusions are summarized in Sec.~\ref{sec:summary}.

\section{Theory}
\label{sec:theory}
This section is divided into two parts: the first describes the physical system and its Hamiltonian, while the second presents the theoretical treatment of the dynamics of the open quantum system and the measurement problem. 

\subsection{Physical system and Hamiltonian}
\label{subsec:system}
 
\begin{figure}[h]
	\centering
	\includegraphics[width=.45\textwidth]{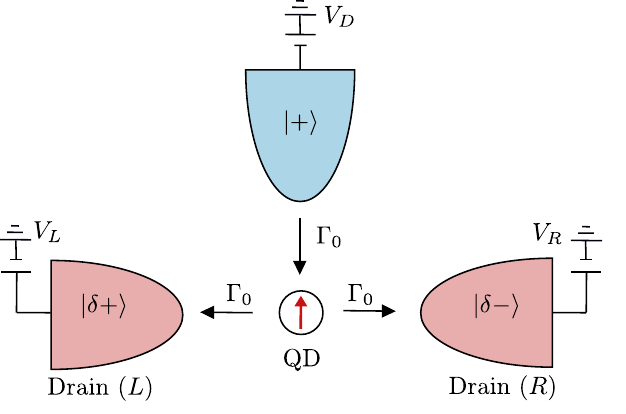}
	\caption{Schematic of the open quantum system considered in this work. A single-level quantum dot is tunnel coupled, with rate $\Gamma_0$, to two ferromagnetic drain electrodes labeled left (L) and right (R). These drain leads perform spin-selective detection, allowing electrons polarized along the $\delta\pm$ directions, with $\delta = x, y, z$, to tunnel out of the quantum dot. The upper lead is included only for illustration purposes and indicates a possible source electrode used to initialize the quantum dot with a spin-polarized electron, prepared in the spin $+$ (i.e., spin $\uparrow$ along the $z$ axis) state.}
	\label{fig_system}
\end{figure}

Figure~\ref{fig_system} illustrates the physical system under investigation: a semiconductor quantum dot coupled to two ferromagnetic leads with opposite spin polarizations (anti-parallel alignment). The drain leads are assumed to be half-metals~\cite{Katsnelson2008}, such that only spin $+$ ($-$) electrons can tunnel into the left (right) lead. The quantization axis for spin detection can be chosen along the $x$, $y$, or $z$ direction.
The system is initialized with an electron in the quantum dot with spin-up state along the $z$ axis, $\ket{+}=\ket{\uparrow}$. The source lead shown in Fig.~\ref{fig_system} is included for illustration only and is not considered in the simulations; its role is to enable on-demand initialization of the spin-up electron in the quantum dot, allowing multiple repetitions of the detection procedure.
After initialization, the electron spin undergoes coherent precession driven by a transverse magnetic field applied along the $x$ axis. As time evolves, the spin state explores different orientations on the Bloch sphere and can eventually tunnel out of the quantum dot into one of the two ferromagnetic drain leads. This tunneling process is inherently stochastic and is simulated by randomly selecting outcomes from a predefined probability distribution generated via the Lindblad formalism, effectively mimicking an experimental detection protocol. These simulated spin-resolved tunneling events form the basis for reconstructing the qubit dynamics.
 
The system Hamiltonian is defined as
\begin{equation}\label{H}
    H =  H_0 + H_{\mathrm{L, \delta}} + H_{\mathrm{R},\delta} + V_\delta,
\end{equation}
where $H_{\mathrm{L},\delta}$ and $H_{\mathrm{R},\delta}$ represent the free-particle Hamiltonians of the left and right drain leads:
\begin{equation}
H_{\eta,\delta} = \sum_{\mathbf{k}} \sum_{\lambda \in \{+,-\}} \varepsilon_{\mathbf{k}\eta\delta\lambda} \, c_{\mathbf{k}\eta\delta\lambda}^\dagger c_{\mathbf{k}\eta\delta\lambda},
\end{equation}
with $\varepsilon_{\mathbf{k}\eta\delta\lambda}$ denoting the energy of an electron characterized by wave vector $\mathbf{k}$ in lead $\eta$ and spin orientation $\lambda$, defined along the axis $\delta \in \{x,y,z\}$. The operators $c_{\mathbf{k}\eta\delta\lambda}$ and $c_{\mathbf{k}\eta\delta\lambda}^\dagger$ annihilate and create, respectively, an electron in lead $\eta$ with wave vector $\mathbf{k}$ and spin alignment $\lambda$ along direction $\delta$.
Additionally, the upper lead shown in Fig.~\ref{fig_system} is magnetized along the $z$ axis with spin orientation $+$. Although not included in the simulations, this lead serves to inject spin-polarized electrons into the quantum dot for initialization, enabling repeated experimental cycles. This configuration ensures spin-selective tunneling along multiple axes, providing access to complementary spin projections and thereby enabling a complete qubit state readout.

For the quantum dot, the Hamiltonian is given by
\begin{equation}
    H_0 = \sum_{\sigma \in \{\uparrow, \downarrow \}} \varepsilon_0 \, d_{\sigma}^{\dagger} d_{\sigma} 
    + \Omega \left( d_{\uparrow}^{\dagger} d_{\downarrow} + d_{\downarrow}^{\dagger} d_{\uparrow} \right),
    \label{H0}
\end{equation}
where $d_\sigma$ ($d_\sigma^\dagger$) annihilates (creates) an electron with spin $\sigma$ in the quantum dot, and $\varepsilon_0$ is the single-particle energy level of the dot. Here, $\uparrow$ (also denoted as $+$) refers to spin up along the $z$ axis, while $\downarrow$ (also denoted as $-$) refers to spin down along the $z$ axis. The parameter $\Omega$ represents the spin-flip coupling rate, induced by an external magnetic field applied along the $x$ direction.

The coupling between the quantum dot and the ferromagnetic leads is described by the interaction Hamiltonian $V_\delta$:
\begin{equation}\label{Vtotal}
V_\delta = \sum_{\mathbf{k}} \sum_{\eta \in \{L, R\}} \sum_{\lambda \in \{+,-\}}
\left( v_{\eta} \, c_{\mathbf{k}\eta\delta\lambda}^\dagger d_{\delta\lambda} + \mathrm{H.c.} \right),
\end{equation}
where $\delta \in \{x,y,z\}$ denotes the magnetization axis of the leads. Both leads are assumed to be ferromagnetic with magnetizations along the same axis but in opposite directions, i.e., they are antiparallel aligned. This configuration ensures spin-selective tunneling, enabling measurements along different spin projections.
Here, $d_{\delta\lambda}$ represents the annihilation operator for an electron in the quantum dot with spin projection $\lambda$ along axis $\delta$. For $\delta = z$, we have $d_{\delta\lambda} = d_{\lambda} = d_\sigma$, where $\lambda = +$ corresponds to spin $\uparrow$ and $\lambda = -$ to spin $\downarrow$. For $\delta = x$ or $y$, the operators are expressed as
\begin{eqnarray}
d_{\phi +} &=& \frac{e^{i\phi/2} d_{\uparrow} + e^{-i\phi/2} d_{\downarrow}}{\sqrt{2}}, \\
d_{\phi -} &=& \frac{-e^{i\phi/2} d_{\uparrow} + e^{-i\phi/2} d_{\downarrow}}{\sqrt{2}},
\end{eqnarray}
where $\phi = 0$ for magnetization along the $x$ axis and $\phi = \pi/2$ for magnetization along the $y$ axis. This representation follows the model introduced by Weiss and K\"onig~\cite{weiss2017} to account for ferromagnetic leads with magnetization orientation $\phi$ in the $x$–$y$ plane.
For simplicity, we assume that the coupling parameters $v_\eta$ are spin- and energy-independent. The interaction Hamiltonian in Eq.~\eqref{Vtotal} provides the basis for deriving the Lindblad master equation, which governs the non-unitary evolution of the quantum dot's reduced density matrix under the influence of the leads.

\subsection{Single-Qubit Model}
\label{subsec:singlequbit}

Following the qubit representation of the fermionic operators $d_\uparrow$ and $d_\downarrow$~\cite{Souza17}, we have
\begin{eqnarray}
\label{dup} d_\uparrow &=& \sigma_{-} \otimes I, \\ 
\label{ddw} d_\downarrow &=& \sigma_z \otimes \sigma_{-},
\end{eqnarray}
where $\sigma_{-} = (\sigma_x - i \sigma_y)/2$, $I$ is the $2 \times 2$ identity matrix, and $\sigma_x$, $\sigma_y$, and $\sigma_z$ are the Pauli matrices. This representation satisfies the fermionic anticommutation relation
$\{ d_\sigma, d_{\sigma'}^\dagger \} = \delta_{\sigma,\sigma'}$.
Based on this mapping, the quantum dot Hamiltonian takes the form
\begin{equation}
H_0 = \Omega
\begin{pmatrix}
0 & 0 & 0 & 0 \\
0 & 0 & -1 & 0 \\
0 & -1 & 0 & 0 \\
0 & 0 & 0 & 0
\end{pmatrix},
\end{equation}
expressed in the basis $\{\ket{00}, \ket{01}, \ket{10}, \ket{11}\}$. For simplicity, we set $\varepsilon_0 = 0$. In an open quantum system, states such as $\ket{00}$ and $\ket{11}$ can be accessed due to particle-number fluctuations induced by the reservoirs. However, for a closed system initialized in $\ket{01}$ (spin-up), the dynamics are restricted to the subset $\{\ket{01}, \ket{10}\}$, as the particle number remains fixed.
To describe this situation, we introduce a compact single-qubit Hamiltonian,
\begin{equation}
h_0 = \Omega
\begin{pmatrix}
0 & -1 \\
-1 & 0
\end{pmatrix}
= -\Omega \sigma_x,
\end{equation}
expressed in the basis $\{\ket{01}, \ket{10}\}$, or equivalently $\{\ket{\uparrow}, \ket{\downarrow}\}$. The time evolution of the state vector is then ($\hbar=1$)
\begin{equation}
\ket{\psi(t)} = \cos(\Omega t)\ket{\uparrow} + i\sin(\Omega t)\ket{\downarrow},
\end{equation}
assuming $\ket{\uparrow}$ as the initial state. The corresponding density matrix $\rho_{2\times 2} = \ket{\psi(t)}\bra{\psi(t)}$ is
\begin{equation}
\rho_{2\times 2} =
\begin{pmatrix}
\cos^2(\Omega t) & -i\cos(\Omega t)\sin(\Omega t) \\
i\cos(\Omega t)\sin(\Omega t) & \sin^2(\Omega t)
\end{pmatrix}.
\end{equation}
In our simulated detection experiment, we aim to recover this density matrix approximately. Specifically, we expect
\begin{eqnarray}
\rho_{2\times 2}^{\uparrow \uparrow} &=& \cos^2(\Omega t), \label{rho22_1} \\ 
\rho_{2\times 2}^{\downarrow \downarrow} &=& \sin^2(\Omega t),  \label{rho22_2}\\ 
\mathrm{Re}\{\rho_{2\times 2}^{\uparrow \downarrow}\} &=& 0, \label{rho22_3} \\ 
\mathrm{Im}\{\rho_{2\times 2}^{\uparrow \downarrow}\} &=& -\cos(\Omega t)\sin(\Omega t). \label{rho22_4}
\end{eqnarray}
To implement the experimental scenario for detecting tunneling events, the system must be treated as open, which we model using the Lindblad formalism, as shown in details in the Appendix~\ref{appendix:openmeas}. After the derivations, we obtain the equation:

\begin{eqnarray}\label{rho_lindblad_theta}
&& \frac{\partial}{\partial \theta} \rho(\theta) = -i\left[\frac{H_0}{\Omega}, \rho(\theta)\right] \\
&& -\frac{1}{2} \frac{\Gamma_{L,\delta,+}}{\Omega} \big[ d_{\delta +}^\dagger d_{\delta +} \rho(\theta) - 2 d_{\delta +} \rho(\theta) d_{\delta +}^\dagger + \rho(\theta) d_{\delta +}^\dagger d_{\delta +} \big] \nonumber \\
&& -\frac{1}{2} \frac{\Gamma_{R,\delta,-}}{\Omega} \big[ d_{\delta -}^\dagger d_{\delta -} \rho(\theta) - 2 d_{\delta -} \rho(\theta) d_{\delta -}^\dagger + \rho(\theta) d_{\delta -}^\dagger d_{\delta -} \big].\nonumber
\end{eqnarray}
Here $\theta = \Omega t/\hbar$ is the dimensionless time (with $\hbar = 1$ throughout), 
$\Gamma_{L,\delta,+} = \Gamma_{R,\delta,-} = \Gamma_0$ are the tunneling 
rates for the left and right ferromagnetic leads respectively, and 
$d_{\delta\pm}$ are the spin-selective annihilation operators defined 
in Sec.~\ref{subsec:singlequbit} for $\delta = x, y$ and 
$d_{\delta\pm} = d_\sigma$ for $\delta = z$, as introduced in 
Eqs.~(\ref{dup})--(\ref{ddw}).

Solving master equation in Eq.~\eqref{rho_lindblad_theta} provides the time evolution of the reduced density matrix $\rho(\theta)$, from which we extract spin-resolved occupation probabilities. These probabilities serve as input for simulating stochastic detection events, generating synthetic measurement outcomes that mimic experimental spin-resolved tunneling from the quantum dot.

\subsection{Reduced Density Matrix}
\label{subsec:stokes}

The density matrix can be represented in the computational basis $\{\ket{00}, \ket{01}, \ket{10}, \ket{11}\}$ as
\begin{equation}
\rho(t) = \sum_{i,j,l,m=0}^1 \rho_{ij,lm}(t) \ket{ij}\bra{lm},
\end{equation}
where $\rho_{00,00}$ denotes the probability of double occupancy, $\rho_{01,01} = \rho_{\uparrow\uparrow}$ and $\rho_{10,10} = \rho_{\downarrow\downarrow}$ correspond to the probabilities of finding a single electron with spin $\ket{\uparrow}$ or $\ket{\downarrow}$ along the $z$ axis, respectively, and $\rho_{11,11}$ represents the empty quantum dot state.
In our model, double occupancy is excluded because we assume that a single electron is initially prepared in the quantum dot and subsequently tunnels into one of the leads. Thus, charge transport occurs unidirectionally from the dot to the lead. This single-electron regime can be realized under Coulomb blockade conditions~\cite{grabert1992}.

Our goal is to reconstruct, via tomographic measurements, the reduced block of the density matrix associated with the odd occupation states $\ket{01}$ and $\ket{10}$, namely
\begin{equation}\label{rho_odd_projectors}
\rho^{\mathrm{o}} = P_{\mathrm{o}} \rho P_{\mathrm{o}},
\end{equation}
where $P_{\mathrm{o}} = \ket{01}\bra{01} + \ket{10}\bra{10} = \ket{\uparrow}\bra{\uparrow} + \ket{\downarrow}\bra{\downarrow}$. This reduced density matrix can be written as
\begin{equation}
\rho^{\mathrm{o}} = \begin{pmatrix} \rho^{\mathrm{o}}_{++} & \rho^{\mathrm{o}}_{+-} \\ \rho^{\mathrm{o}}_{-+} & \rho^{\mathrm{o}}_{--} \end{pmatrix}.
\label{rho_reduced}
\end{equation}

The diagonal elements $\rho^{\mathrm{o}}_{++}$ and $\rho^{\mathrm{o}}_{--}$ can be directly obtained from spin-polarized measurements along the $z$ axis using ferromagnetic leads. However, the off-diagonal elements $\rho^{\mathrm{o}}_{+-}$ and $\rho^{\mathrm{o}}_{-+}$ require measurements along the $x$ and $y$ spin projections. Specifically, we have
\begin{align}
\rho^{\mathrm{o}}_{x+ x+} &= \frac{1}{2} \left( \rho^{\mathrm{o}}_{++} + \rho^{\mathrm{o}}_{+-} + \rho^{\mathrm{o}}_{-+} + \rho^{\mathrm{o}}_{--} \right), \label{rhoxpxp} \\
\rho^{\mathrm{o}}_{x- x-} &= \frac{1}{2} \left( \rho^{\mathrm{o}}_{++} - \rho^{\mathrm{o}}_{+-} - \rho^{\mathrm{o}}_{-+} + \rho^{\mathrm{o}}_{--} \right), \label{rhoxmxm}
\end{align}
and
\begin{align}
\rho^{\mathrm{o}}_{y+ y+} &= \frac{1}{2} \left( \rho^{\mathrm{o}}_{++} + i \rho^{\mathrm{o}}_{+-} - i \rho^{\mathrm{o}}_{-+} + \rho^{\mathrm{o}}_{--} \right), \label{rhoypyp} \\
\rho^{\mathrm{o}}_{y- y-} &= \frac{1}{2} \left( \rho^{\mathrm{o}}_{++} - i \rho^{\mathrm{o}}_{+-} + i \rho^{\mathrm{o}}_{-+} + \rho^{\mathrm{o}}_{--} \right). \label{rhoymym}
\end{align}

From Eqs.~(\ref{rhoxpxp}) and (\ref{rhoypyp}), the real and imaginary parts of $\rho^{\mathrm{o}}_{+-}$ can be extracted as
\begin{align}
\mathrm{Re}\{\rho^{\mathrm{o}}_{+-}\} &= \rho^{\mathrm{o}}_{x+ x+} - \frac{1}{2} \left( \rho^{\mathrm{o}}_{++} + \rho^{\mathrm{o}}_{--} \right), \label{Rerhopm} \\
\mathrm{Im}\{\rho^{\mathrm{o}}_{+-}\} &= - \rho^{\mathrm{o}}_{y+ y+} + \frac{1}{2} \left( \rho^{\mathrm{o}}_{++} + \rho^{\mathrm{o}}_{--} \right). \label{Imrhopm}
\end{align}
Alternatively, Eqs.~(\ref{rhoxmxm}) and (\ref{rhoymym}) provide equivalent expressions based on the minus spin components.
In summary, reconstructing the reduced density matrix $\rho^{\mathrm{o}}$ requires determining only four quantities: $\rho^{\mathrm{o}}_{++}$, $\rho^{\mathrm{o}}_{--}$, $\rho^{\mathrm{o}}_{x+ x+}$, and $\rho^{\mathrm{o}}_{y+ y+}$.

\section{Stochastic Simulation of Quantum Dot Qubit Tomography}
\label{sec:simulations}

In this section, we introduce a stochastic framework to simulate experimentally realistic conditions for the characterization of a quantum dot qubit, where spin-resolved tunneling events are sequentially recorded over time. The detection process is modeled by randomly sampling tunneling events according to probabilities derived from the time-dependent density matrix $\rho(t)$.

Using Eq.~\eqref{rho_odd_projectors}, we obtain the reduced density matrix $\rho^{\mathrm{o}}$, from which we extract the components $\rho^{\mathrm{o}}_{++}$, $\rho^{\mathrm{o}}_{--}$, as well as $\rho^{\mathrm{o}}_{x+ x+}$ and $\rho^{\mathrm{o}}_{y+ y+}$ via Eqs.~\eqref{rhoxpxp} and \eqref{rhoypyp}. The tunneling probabilities associated with spin projections along the $x$, $y$, and $z$ directions are then given by
\begin{eqnarray}
p_{x+}(t) &=& \Delta \Gamma_0 \, \rho^{\mathrm{o}}_{x+ x+}(t),\label{pxp} \\
p_{y+}(t) &=& \Delta \Gamma_0 \, \rho^{\mathrm{o}}_{y+ y+}(t),\label{pyp} \\
p_{z+}(t) &=& \Delta \Gamma_0 \, \rho^{\mathrm{o}}_{++}(t),\label{pzp} \\
p_{z-}(t) &=& \Delta \Gamma_0 \, \rho^{\mathrm{o}}_{--}(t),\label{pzm}
\end{eqnarray}
where $\Gamma_0$ is a constant tunneling rate, assumed equal for all ferromagnetic leads for simplicity, and $\Delta$ is the temporal resolution of the measurement window centered around time $t$. These probabilities are then used to generate a list of detection times and corresponding spin outcomes by sampling from a time-resolved weighted distribution. This approach mimics the stochastic nature of tunneling events observed in real experiments and enables the reconstruction of the qubit state via statistical averaging over many simulated detection events.

Determining the tunneling probabilities $p_{\delta\sigma}$ requires repeated initialization and measurement of the quantum system over $R$ independent trials. For example, excitation of the quantum dot can be achieved using a circularly polarized laser pulse, which generates a single electron prepared in a well-defined spin state within the confining potential~\cite{fujita2019,villasboas2007}. Following initialization, the electron eventually tunnels into an adjacent lead, a process that can be resolved using charge detection techniques such as single-electron current or charge sensing~\cite{kurzmann2019}. By recording the timing and spin-dependent outcomes of individual tunneling events, a statistically significant data set is obtained. These stochastically generated tunneling statistics are subsequently analyzed using supervised machine-learning techniques, which infer the underlying density matrix elements from the measured probability distributions and enable reconstruction of the qubit state.

In our simulations, we assume continuous monitoring of the system over a total duration $T$, which is discretized into $M$ subintervals of width $\Delta$, such that $T = M\Delta$. The measurement protocol is repeated $R$ times, yielding the following empirical probability:

\begin{equation}\label{stochastic_probability}
\Pi_{\lambda \sigma}(i) = \frac{N_{\lambda \sigma}(i)}{R},
\end{equation}
where $N_{\lambda \sigma}(i)$ denotes the number of detected tunneling events within the $i$-th time interval, associated with the spin-polarized lead labeled by $\lambda \sigma$. For sufficiently large $R$, we expect the empirical probability to converge to the theoretical one, i.e., $\Pi_{\lambda \sigma}(i) \to p_{\lambda \sigma}(i)$.

To simulate the detection process, we define a set of integers $\mathcal{S} = \{1, 2, 3, \dots, M, M+1\}$. In each trial, an integer $i \in \mathcal{S}$ is randomly selected according to the probability distribution $p_{\lambda \sigma}(i)$, where $i \in [1, M]$ corresponds to a detection event occurring in the $i$-th time interval. The additional element $M+1$ in $\mathcal{S}$ accounts for the possibility that no detection occurs during the observation window, with probability
\begin{equation}
q_{\lambda \sigma}(M+1) = 1 - \sum_{i=1}^{M} p_{\lambda \sigma}(i).
\end{equation}
This framework allows us to generate detection events that statistically reflect the underlying quantum dynamics of the system, as governed by the Lindblad equation~\eqref{rho_lindblad_theta}. After performing $R$ trials, we obtain a distribution of detection events represented by $N_{\lambda \sigma}(i)$.

From this counting, we construct the empirical probability distribution using Eq.~\eqref{stochastic_probability}, which serves as a stochastic estimate of the underlying quantum probabilities. Furthermore, by dividing this distribution by the factor $\Delta \Gamma_0$, we recover the four spin-resolved density matrix elements described in Eqs.~\eqref{pxp}--\eqref{pzm}, i.e.,
\begin{eqnarray}
\rho^{\mathrm{o}}_{x+ x+}(t) &\approx& \Pi_{x+}(t) / (\Delta \Gamma_0 ), \label{pxp_approx} \\
\rho^{\mathrm{o}}_{y+ y+}(t)  &\approx& \Pi_{y+}(t) / (\Delta \Gamma_0 ),\label{pyp_approx} \\
\rho^{\mathrm{o}}_{++}(t) &\approx& \Pi_{z+}(t) / (\Delta \Gamma_0 ),\label{pzp_approx} \\
\rho^{\mathrm{o}}_{--}(t) &\approx& \Pi_{z-}(t) / (\Delta \Gamma_0 ),\label{pzm_approx}
\end{eqnarray}
thereby linking the simulated detection statistics directly to the quantum state of the system.
Overall, this simulation framework offers a flexible and experimentally grounded platform for testing quantum dot qubit tomography protocols.

\subsection{Algorithm Description}
\label{subsec:algorithm}

Our framework consists of repeatedly performing the experiment by initializing 
a single electron with spin-up (along the $z$ axis) in the quantum dot and 
subsequently detecting it as a current flowing into a ferromagnetic lead whose 
magnetization is tuned to collect a specific spin orientation. As the experiment 
is repeated, detection events occur more frequently at certain times than at 
others due to the temporal evolution of the qubit inside the quantum dot, which 
undergoes Rabi oscillations. In addition, the stochastic nature of spin-polarized 
tunneling events introduces randomness, where some outcomes may never be observed 
within a finite number of trials, similar to rolling dice in a game.

\begin{figure}[t]
	\centering
	\includegraphics[width=\columnwidth]{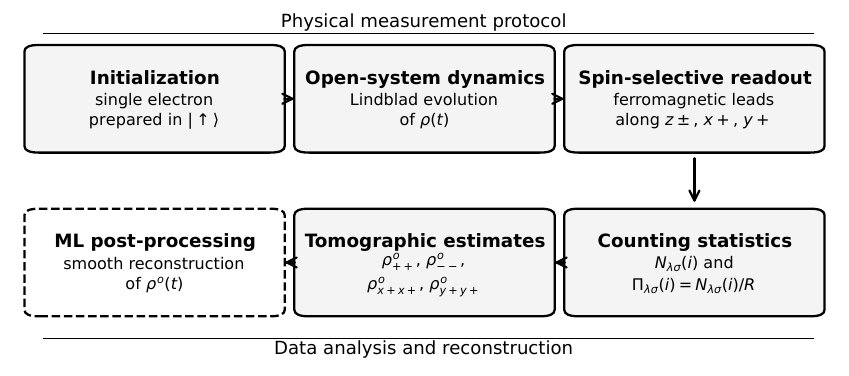}
	\caption{Workflow of the proposed spin-polarized transport tomography protocol. 
	The upper row illustrates the physical measurement stage: after initialization 
	of a single electron in the spin-up state $|\!\uparrow\rangle$ in the quantum 
	dot, the spin evolves according to the open-system Lindblad dynamics and is 
	detected through spin-selective tunneling into ferromagnetic drain leads with 
	different magnetization directions. Repeated measurements generate stochastic 
	counting data $N_{\lambda\sigma}(i)$, from which the empirical probabilities 
	$\Pi_{\lambda\sigma}(i) = N_{\lambda\sigma}(i)/R$ are extracted. The lower row 
	illustrates the data analysis and reconstruction stage: these probabilities 
	provide estimates of the density-matrix elements 
	$\rho^{\mathrm{o}}_{++}$, $\rho^{\mathrm{o}}_{--}$, 
	$\rho^{\mathrm{o}}_{x+x+}$, and $\rho^{\mathrm{o}}_{y+y+}$, 
	which are used as input to a supervised Gaussian Process Regression (GPR) 
	model that reconstructs smooth time-dependent density-matrix elements 
	$\rho^{\mathrm{o}}(t)$ from the simulated counting statistics, including 
	the off-diagonal coherences.}
	\label{fig:workflow}
\end{figure}

The complete workflow of the proposed protocol is illustrated in 
Fig.~\ref{fig:workflow}. The protocol proceeds in two stages. In the 
\textit{physical measurement stage}, the system is initialized in the spin-up 
state $|\!\uparrow\rangle$, evolves under the open-system Lindblad dynamics, 
and is detected through spin-selective tunneling into ferromagnetic leads along 
the $z\pm$, $x+$, and $y+$ directions. Repeated measurements generate stochastic 
counting data $N_{\lambda\sigma}(i)$, from which the empirical probabilities 
$\Pi_{\lambda\sigma}(i) = N_{\lambda\sigma}(i)/R$ are extracted. In the 
\textit{data analysis and reconstruction stage}, these probabilities provide 
estimates of the density-matrix elements 
$\rho^{\mathrm{o}}_{++}$, $\rho^{\mathrm{o}}_{--}$, 
$\rho^{\mathrm{o}}_{x+x+}$, and $\rho^{\mathrm{o}}_{y+y+}$, 
which are then fed into a supervised Gaussian Process Regression (GPR)~\cite{rasmussen2006} model 
that reconstructs the smooth time-dependent reduced density matrix $\rho^{\mathrm{o}}(t)$, 
including the off-diagonal coherences, at any time.

The GPR model is trained on the stochastic density-matrix estimates
\[
\bigl\{\rho^{\mathrm{o}}_{++}(\theta_i),\, \rho^{\mathrm{o}}_{--}(\theta_i),\,
\rho^{\mathrm{o}}_{x+x+}(\theta_i),\, \rho^{\mathrm{o}}_{y+y+}(\theta_i)\bigr\}
\]
at the detected time points $\theta_i$, where the input features are the
detection times and the target outputs are the corresponding density-matrix
element values. The model uses a composite kernel consisting of a
squared-exponential (RBF) term, which captures the smooth oscillatory
behavior of the Rabi dynamics, and a WhiteKernel term, which models the
stochastic noise in the counting data. The kernel hyperparameters (length
scale and noise level) are optimized automatically via maximization of the
log marginal likelihood, using multiple random restarts ($n_{\mathrm{restarts}}
= 10$) to avoid local optima. The trained model predicts the full continuous
time evolution over $\theta \in [0, 2\pi]$, providing both a mean prediction
and quantified $\pm 1\sigma$ uncertainty bands that reflect the statistical
quality of the detection data at each time point.

For comparison, a direct inversion of the detected data 
via linear interpolation --- without any ML post-processing --- yields 
significantly noisier reconstructions, demonstrating that the GPR 
step is essential for a reliable density-matrix reconstruction 
from sparse and irregularly sampled stochastic data.

\section{Simulated Detection Results}
\label{sec:randomdata}

\begin{figure}[tb]
	\centering\includegraphics[width=0.95\linewidth]{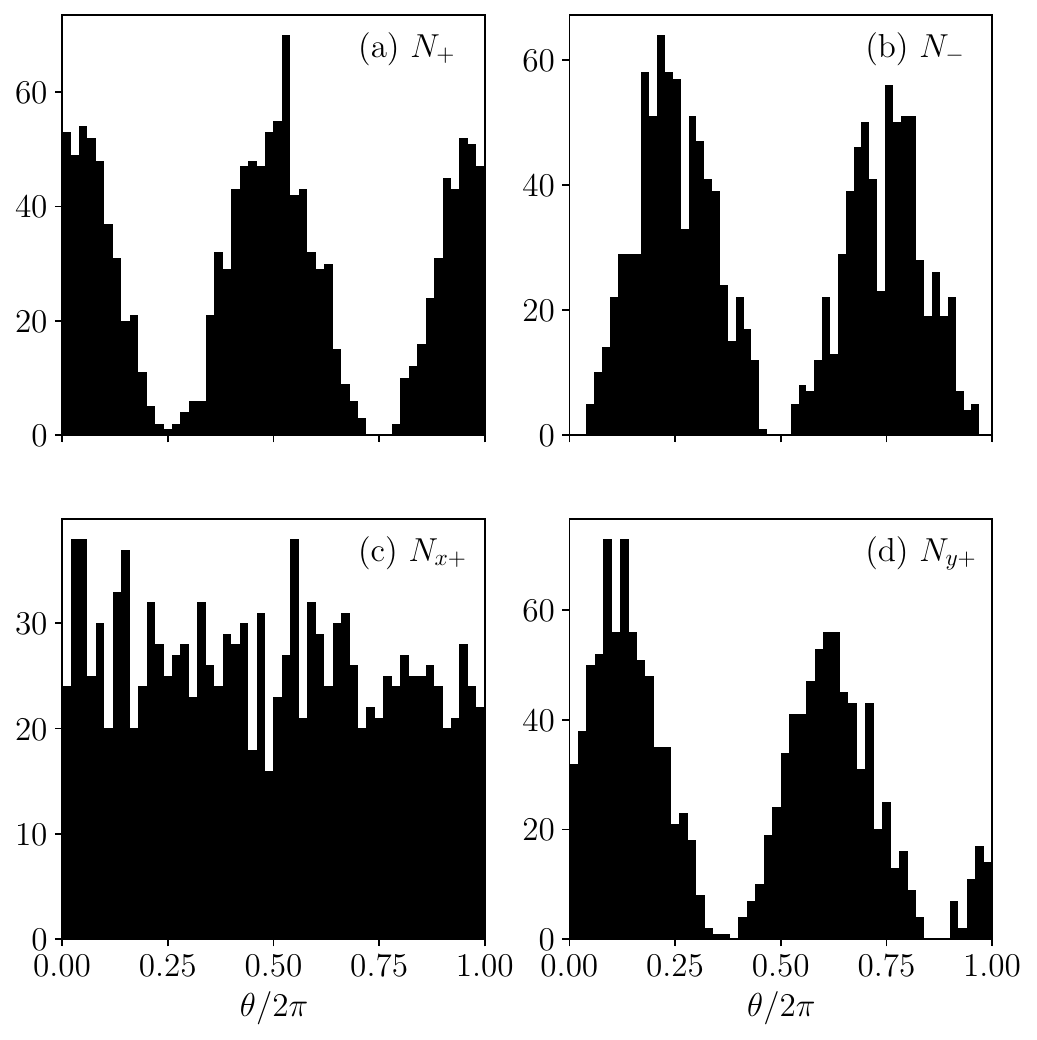}
	\caption{Simulated spin-resolved tunneling event counts as a function 
of time $\theta/2\pi$, for four ferromagnetic lead alignments: 
(a)~$z+$, (b)~$z-$, (c)~$x+$, and (d)~$y+$. 
Panels (a), (b), and (d) display characteristic clusters reflecting 
the Rabi oscillations of the qubit spin precessing in the $y$--$z$ 
plane of the Bloch sphere under the transverse magnetic field. 
The clusters gradually diminish over time due to decoherence induced 
by coupling to the drain leads. Panel (c) shows a purely exponential 
decay without oscillatory features, as the spin projection along 
the $x$ direction remains constant under the transverse field and 
does not undergo precession.
Parameters: $\Omega = 1~\mu\mathrm{eV}$ ($242~\mathrm{MHz}$), 
$\Gamma_0 = 0.05\,\Omega = 0.05~\mu\mathrm{eV}$ ($12.1~\mathrm{MHz}$), 
and $R = 10{,}000$.}
	\label{fig_counting_data}
\end{figure}

\begin{figure}[tb]
    \centering
    \includegraphics[width=0.95\linewidth]{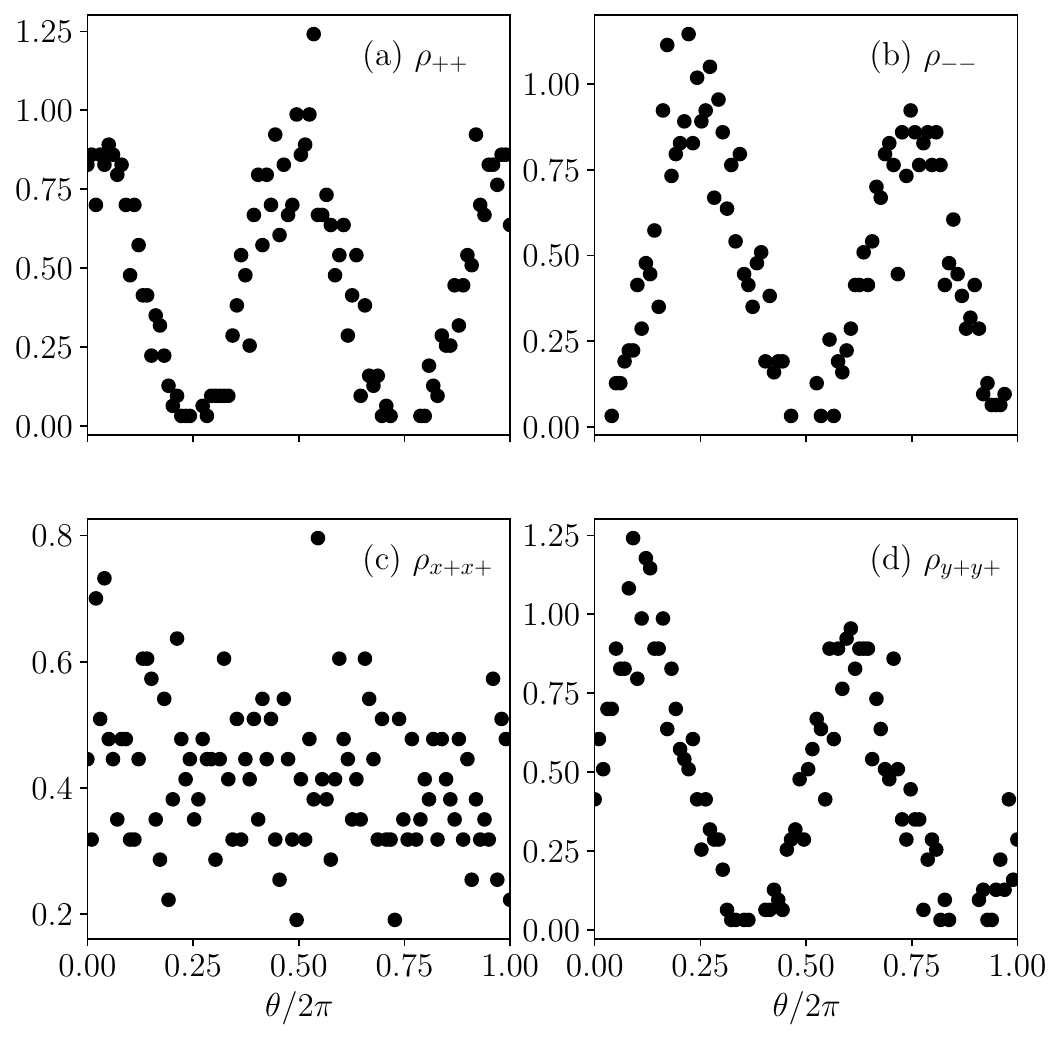}
    \caption{Stochastic estimates of the spin-resolved density-matrix 
elements $\rho^{\mathrm{o}}_{++}$, $\rho^{\mathrm{o}}_{--}$, 
$\rho^{\mathrm{o}}_{x+x+}$, and $\rho^{\mathrm{o}}_{y+y+}$, 
obtained from the tunneling event counts of 
Fig.~\ref{fig_counting_data} via Eqs.~(\ref{pxp_approx})--(\ref{pzm_approx}). 
These four quantities serve as input to the GPR model and allow 
determination of the full off-diagonal coherence via 
Eqs.~(\ref{Rerhopm})--(\ref{Imrhopm}).
Parameters: $\Omega = 1~\mu\mathrm{eV}$ ($242~\mathrm{MHz}$), 
$\Gamma_0 = 0.05\,\Omega = 0.05~\mu\mathrm{eV}$ ($12.1~\mathrm{MHz}$), 
and $R = 10{,}000$.}
    \label{fig_rhos_from_counting}
\end{figure}

Figure~\ref{fig_counting_data} presents the simulated tunneling events recorded in a lead in four different ferromagnetic alignments: (a) $\ket{+}$ alignment in $z$ direction, (b)  $\ket{-}$ alignment in $z$ direction, (c)  $\ket{x+}$ alignment in $x$ direction and (d)  $\ket{y+}$ alignment in $y$ direction. The $\ket{+}$ component initially exhibits a high count due to the system being initialized in the spin-up state. As time progresses, the $\ket{+}$, $\ket{-}$, and $\ket{y+}$ channels display clusters of tunneling events, as shown in panels (a), (b) and (d). These clusters reflect the characteristic Rabi oscillations of a two-level system, arising from spin precession in the $y$–$z$ plane of the Bloch sphere under the influence of a transverse magnetic field oriented along the $x$ direction, as described by the system Hamiltonian in Eq.~\eqref{H0}.
Over time, all tunneling event clusters gradually diminish due to decoherence effects induced by the coupling to the drain lead, thus reflecting the destructive nature of the detection procedure. In contrast, the tunneling events associated with the $\ket{x+}$ spin polarization, shown in panel (d), exhibit a purely exponential decay without oscillatory features. This behavior is expected, as the spin projection along the $x$ direction remains constant under the transverse field and does not undergo precession. The observed decay thus reflects the gradual depletion of the spin population due to tunneling into the leads.
Using the number of tunneling events $N_{\lambda \sigma}(i)$ shown in Fig.~\ref{fig_counting_data}, we compute the empirical probabilities $\Pi_{\lambda \sigma}(i)$ for each of the four spin components, as defined in Eq.~\eqref{stochastic_probability}. Substituting these probabilities into Eqs.~(\ref{pxp_approx})–(\ref{pzm_approx}), we obtain the approximate values for the corresponding spin-resolved density matrix diagonal elements  $\rho^{\mathrm{o}}_{++}$, $\rho^{\mathrm{o}}_{--}$, $\rho^{\mathrm{o}}_{x+ x+}$, and $\rho^{\mathrm{o}}_{y+ y+}$, presented in Fig.~\ref{fig_rhos_from_counting}. The overall behavior of the four components follows the counting events in Fig.~\ref{fig_counting_data}.

\begin{figure}[tb]
	\centering
	\includegraphics[width=0.95\linewidth]{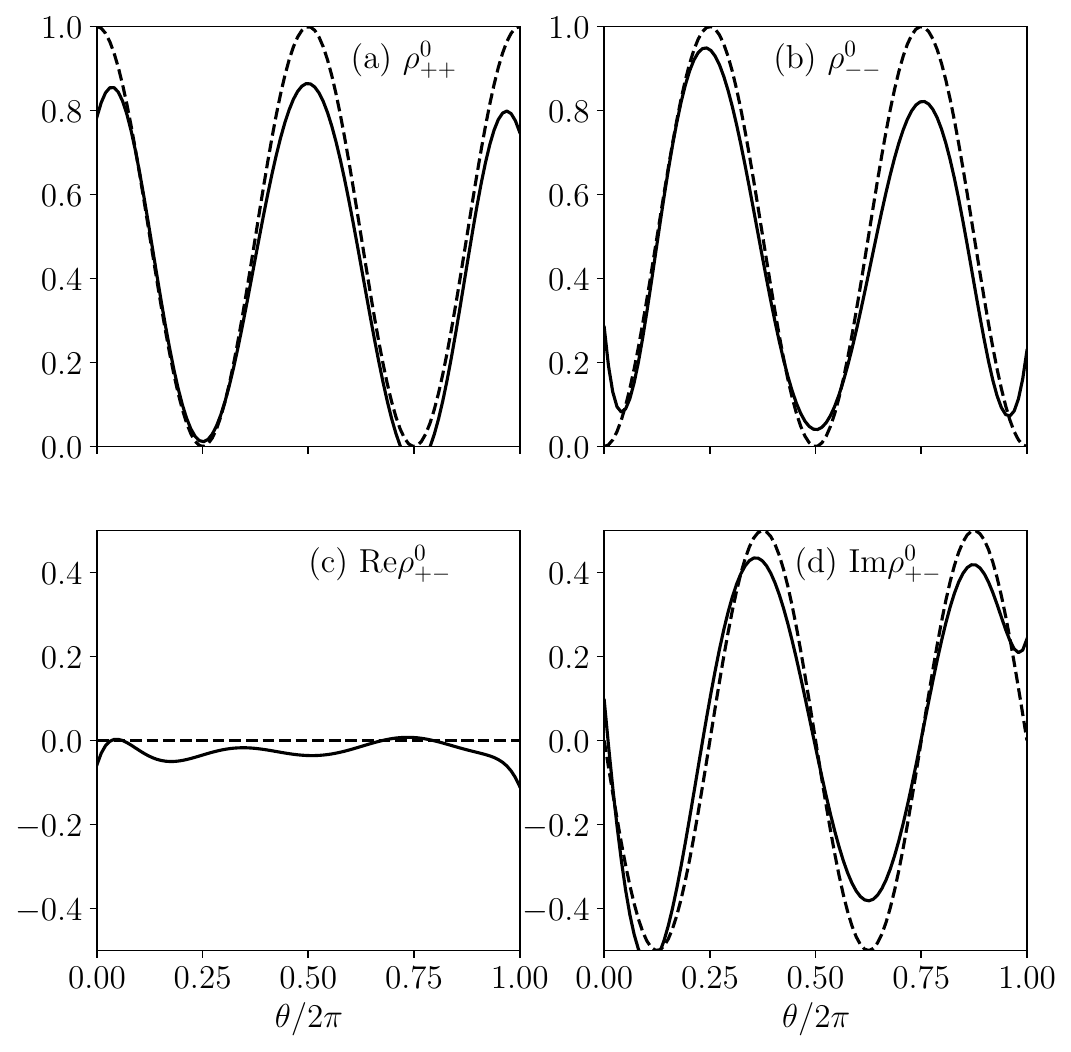}
	\caption{Reconstructed reduced density-matrix elements of 
Eq.~(\ref{rho_reduced}), obtained from the GPR model trained on 
the stochastic estimates of Fig.~\ref{fig_rhos_from_counting}. 
Solid lines show the GPR mean prediction; dashed lines show the 
corresponding evolution for the closed system (no coupling to the 
leads). Panels (a) and (b) display the diagonal elements 
$\rho^{\mathrm{o}}_{++}$ and $\rho^{\mathrm{o}}_{--}$, which exhibit 
damped Rabi oscillations due to the transverse magnetic field and 
coupling to the leads. Panels (c) and (d) display the real and 
imaginary parts of the off-diagonal coherence $\rho^{\mathrm{o}}_{+-}$: 
the real part remains approximately zero by symmetry, while the 
imaginary part oscillates and gradually decays, reflecting the loss 
of quantum coherence induced by the ferromagnetic leads.
Parameters: $\Omega = 1~\mu\mathrm{eV}$ ($242~\mathrm{MHz}$), 
$\Gamma_0 = 0.05\,\Omega = 0.05~\mu\mathrm{eV}$ ($12.1~\mathrm{MHz}$), 
and $R = 10{,}000$.}
	\label{fig_rhos_reconstructed_with_gpr}
\end{figure}

To reconstruct the full-time evolution of these elements, rather than relying on a few sampled values from the stochastic detection process, we employ a supervised machine-learning model~\cite{pedregosa2011}. This model uses the estimated density matrix elements as input and predicts their continuous evolution over time. With the predicted model for $\rho^{\mathrm{o}}_{++}$, $\rho^{\mathrm{o}}_{--}$, $\rho^{\mathrm{o}}_{x+ x+}$, and $\rho^{\mathrm{o}}_{y+ y+}$we reconstruct the complete density matrix
$\rho^{\mathrm{o}}_{++}$, $\rho^{\mathrm{o}}_{--}$ and $\rho^{\mathrm{o}}_{+-}$ via Eqs. (\ref{Rerhopm})-(\ref{Imrhopm}), which are shown in Fig.~\ref{fig_rhos_reconstructed_with_gpr}. The solid lines in Fig.~\ref{fig_rhos_reconstructed_with_gpr} represent the machine-learning predictions, which capture the main features of the randomly generated data while smoothing out statistical fluctuations.
Panels (a) and (b) display the diagonal elements $\rho^{\mathrm{o}}_{++}$ and $\rho^{\mathrm{o}}_{--}$, respectively. Since the system is initialized in the spin-up state, $\rho^{\mathrm{o}}_{++}$ starts near unity. As time evolves, both diagonal elements exhibit damped Rabi oscillations due to the transverse magnetic field and coupling to the leads. Panels (c) and (d) present the real [Fig.~\ref{fig_rhos_reconstructed_with_gpr}(c)] and imaginary [Fig.~\ref{fig_rhos_reconstructed_with_gpr}(d)] parts of the off-diagonal element $\rho^{\mathrm{o}}_{+-}$. The real part remains approximately constant near zero, whereas the imaginary part oscillates and gradually decays to zero. This decay reflects the expected loss of coherence in the system due to coupling with the drain leads. For comparison, the dashed lines represent the corresponding density matrix elements of the closed system (i.e., without coupling to the leads) given by Eqs. (\ref{rho22_1})-(\ref{rho22_4}).

\subsection{Experimental Feasibility}

The coupling parameter $\Omega$ appearing in Eq.~(\ref{H0}) is given by
\begin{equation}
\Omega = h f_{\mathrm{Rabi}} = g \mu_B B_{1},
\end{equation}
where $g$ is the effective $g$-factor of the semiconductor material, $\mu_B = 57.9~\mu\mathrm{eV/T}$ is the Bohr magneton, and $B_{1}=B_{\mathrm{ac}}/2$, with $B_{\mathrm{ac}}$ denoting the amplitude of a radio-frequency (RF) modulated transverse magnetic field. In the present setup, a purely static transverse field is sufficient, as no external magnetic field component along the $z$ direction is considered and consequently no Zeeman splitting is accounted for. Using experimentally reported parameters for GaAs quantum dots, we take $\Omega \approx 1~\mu\mathrm{eV}$, which corresponds to a spin precession (Rabi) frequency of approximately $243~\mathrm{MHz}$. This value is compatible with a transverse field amplitude $B_1$ of about $40~\mathrm{mT}$.\cite{petta2005, koppens2006} 
The tunnel coupling between the quantum dot and the electronic reservoirs, characterized by the rate $\Gamma_0$, sets the timescale for electron escape and detection. In our model, we choose $\Gamma_0 = 0.05\,\Omega = 0.05~\mu\mathrm{eV}$. This choice satisfies the experimentally relevant condition $1 \mu s \gg \Gamma_0^{-1}$,\cite{koppens2006} corresponding to $\Gamma_0 \gg 0.004~\mu\mathrm{eV}$, which lies well within the tunable range of typical quantum-dot devices. 
Experimentally, the coupling strength $\Gamma_0$ can be adjusted with high precision via gate voltages that control the tunnel barrier transparency. As a result, the parameter regime employed in our simulations is experimentally accessible and compatible with state-of-the-art spin-resolved charge detection techniques.

\subsection{Robustness Analysis}

To assess the robustness of the proposed tomography protocol, we investigate 
how the quality of the density-matrix reconstruction depends on the number 
of detection repetitions $R$. Since each run of the stochastic detection 
simulator generates an independent random realization of the tunneling events, 
a single realization is insufficient to reliably quantify the systematic effect 
of $R$ on the reconstruction quality, the run-to-run statistical fluctuation 
can easily mask or even reverse the expected trend. To address this, for each 
value of $R$ we perform $N = 10$ independent realizations of the full pipeline 
(stochastic detection $\rightarrow$ GPR reconstruction), compute the state 
fidelity
\begin{equation}
F(\rho, \sigma) = \left(\mathrm{Tr}\sqrt{\sqrt{\rho}\,\sigma\,\sqrt{\rho}}\right)^2
\label{eq:fidelity}
\end{equation}
of each reconstructed density matrix $\rho$ with respect to the 
closed-system analytical solution $\sigma$, and report the mean fidelity $\bar{F}$ 
and its standard deviation $\sigma_F$ over the $N$ realizations.

\begin{figure}[tb]
\centering
\includegraphics[width=0.95\linewidth]{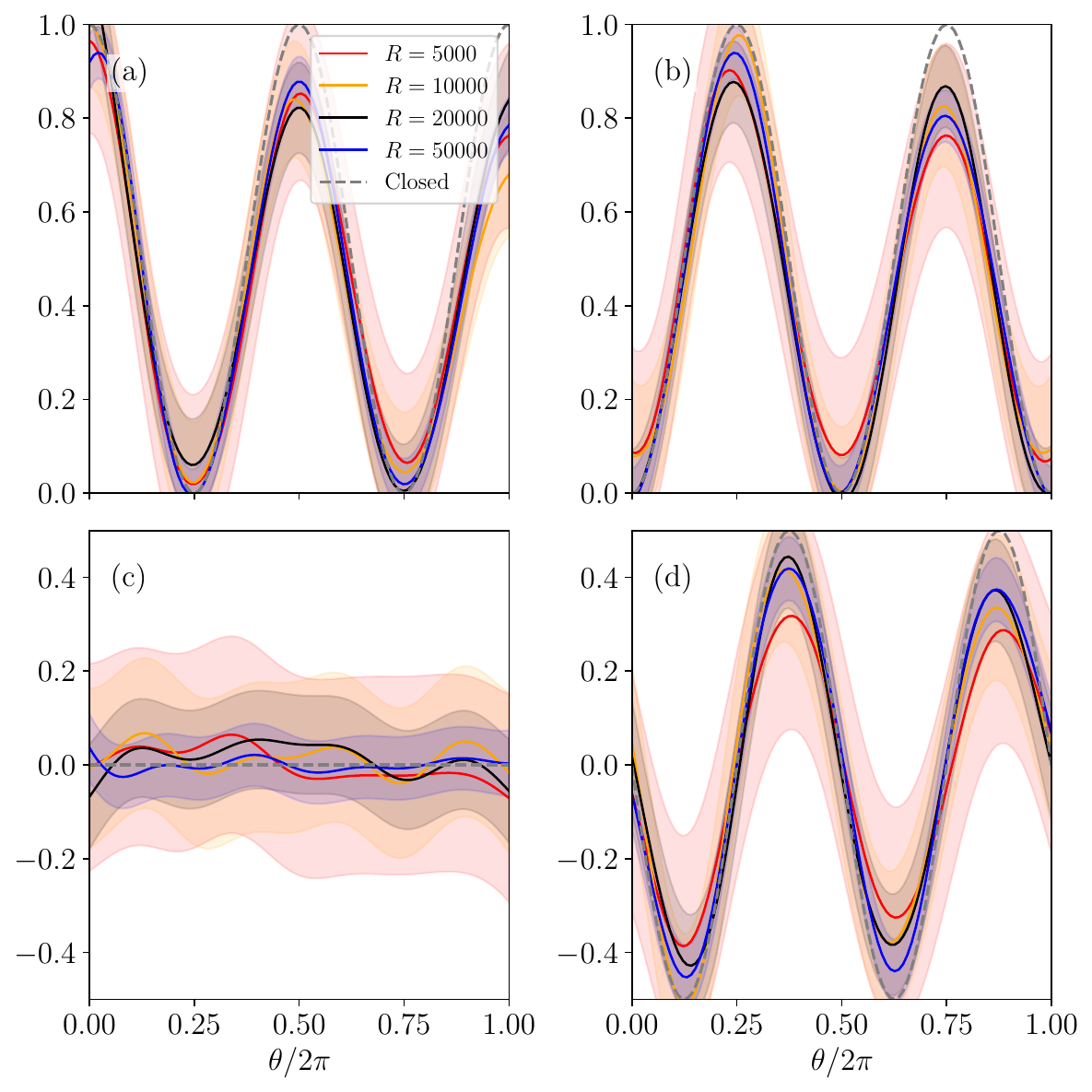}
\caption{GPR-reconstructed density-matrix elements of Eq.~(\ref{rho_reduced}) 
for increasing numbers of detection repetitions $R = 5{,}000$, $10000$, $20000$, 
and $50000$ (red, orange, black, and blue solid lines, respectively), with 
$\pm 1\sigma$ GPR uncertainty bands. The gray dashed line represents the 
corresponding evolution for the closed system (no coupling to the leads). 
Panels (a) and (b) show the diagonal elements $\rho^{\mathrm{o}}_{++}$ and 
$\rho^{\mathrm{o}}_{--}$, whereas panels (c) and (d) display the real and 
imaginary parts of the off-diagonal coherence $\rho^{\mathrm{o}}_{+-}$. 
As $R$ increases, the reconstructed curves converge toward the exact 
open-system dynamics and the uncertainty bands narrow, reflecting the 
improved statistical quality of the detection data.
Parameters: $\Omega = 1~\mu\mathrm{eV}$ ($242~\mathrm{MHz}$) and 
$\Gamma_0 = 0.05\,\Omega = 0.05~\mu\mathrm{eV}$ ($12.1~\mathrm{MHz}$).}
\label{fig_robustness_R_rho}
\end{figure}

\begin{figure}[tb]
\centering
\includegraphics[width=0.95\linewidth]{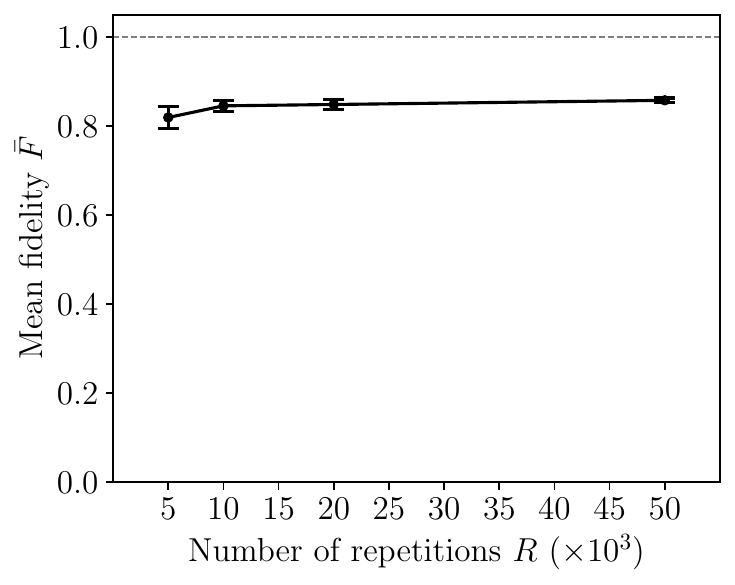}
\caption{Mean state fidelity $\bar{F}$ as a function of the number of 
detection repetitions $R$, computed over $N = 10$ independent stochastic 
realizations at each value of $R$. Error bars represent the run-to-run 
standard deviation $\sigma_F$. The dashed line indicates perfect fidelity 
$\bar{F} = 1$. A clear monotonic increase of $\bar{F}$ with $R$ is observed, 
accompanied by a systematic reduction of $\sigma_F$ from $0.019$ at 
$R = 5\times10^3$ to $0.005$ at $R = 50\times10^3$ (error bars 
represent $\sigma_F$; bars may be smaller than the symbol size for 
large $R$), demonstrating that 
both the reconstruction accuracy and its reproducibility improve with the 
number of experimental repetitions. The residual gap between $\bar{F}$ and 
unity reflects a systematic limitation of the GPR reconstruction associated 
with the $x+$ detection channel, whose spin projection remains constant 
under the transverse field and therefore exhibits low signal-to-noise ratio 
in the stochastic counting data.
Parameters: $\Omega = 1~\mu\mathrm{eV}$ ($242~\mathrm{MHz}$) and 
$\Gamma_0 = 0.05\,\Omega = 0.05~\mu\mathrm{eV}$ ($12.1~\mathrm{MHz}$),
$p = 1$.}
\label{fig_robustness_R_fidelity}
\end{figure}

Figure~\ref{fig_robustness_R_rho} shows the GPR-reconstructed density-matrix elements for 
$R = 5{,}000$, $R = 10{,}000$, $R = 20{,}000$, and $R = 50{,}000$, along with the 
$\pm 1\sigma$ GPR uncertainty bands and the closed-system analytical solution 
(gray dashed line) for reference. As $R$ increases, the reconstructed curves 
converge toward the open-system dynamics and the uncertainty bands 
narrow, reflecting the improved statistical quality of the detection data. 
The corresponding mean fidelities are summarized in Fig.~\ref{fig_robustness_R_fidelity}, where 
a clear monotonic increase of $\bar{F}$ with $R$ is observed, accompanied 
by a systematic reduction of $\sigma_F$ from $0.019$ at $R = 5{,}000$ to 
$0.005$ at $R = 50{,}000$. This demonstrates that the reconstruction accuracy 
and reproducibility of the protocol are directly controlled by the number 
of experimental repetitions, which is an experimentally accessible parameter 
(total acquisition time), and not by intrinsic limitations of the ferromagnetic 
lead configuration.

\begin{figure}[tb]
\centering
\includegraphics[width=0.95\linewidth]{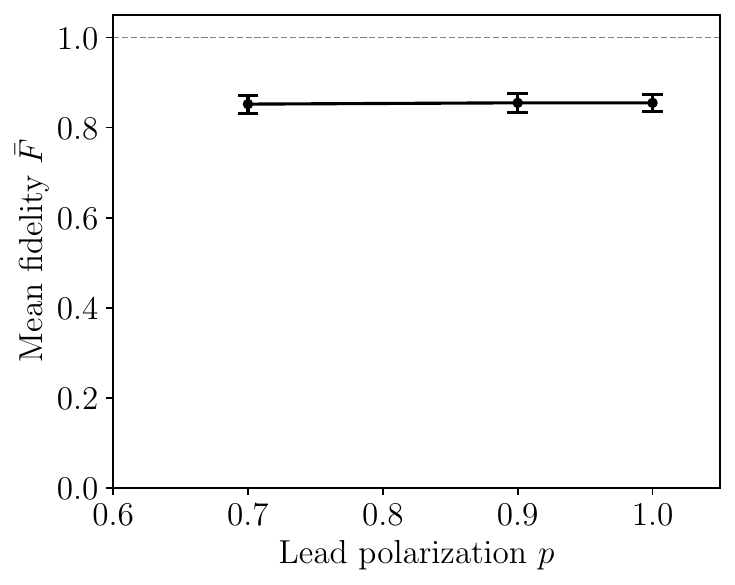}
\caption{Mean state fidelity $\bar{F}$ as a function of the lead spin 
polarization $p$, computed over $N = 10$ independent stochastic realizations 
at each value of $p$. Error bars represent the run-to-run standard deviation 
$\sigma_F \approx 0.012$; bars may be smaller than the symbol size. 
The dashed line indicates perfect fidelity $\bar{F} = 1$. 
The three data points at $p = 0.7$, $0.9$, and $1.0$ yield 
$\bar{F} \approx 0.845$ in all cases, consistent with the analytical result 
that the Lindblad master equation governing the qubit dynamics is exactly 
invariant under $p$ in the antiparallel lead configuration: the 
correct-spin tunneling rate $p\,\Gamma_0$ and the wrong-spin leakage rate 
$(1-p)\,\Gamma_0$ from the opposite lead always sum to $\Gamma_0$, 
leaving the total dissipation rate per spin channel unchanged. 
The protocol is therefore intrinsically robust against finite lead 
spin polarization.
Parameters: $\Omega = 1~\mu\mathrm{eV}$ ($242~\mathrm{MHz}$) and 
$\Gamma_0 = 0.05\,\Omega = 0.05~\mu\mathrm{eV}$ ($12.1~\mathrm{MHz}$),
$R = 10{,}000$.}
\label{fig_fidelity_vs_polarization}
\end{figure}

Furthermore, we investigated the effect of finite lead spin polarization $p$ 
on the reconstruction fidelity. To make this result precise, we show 
analytically that the Lindblad master equation~(\ref{rho_lindblad_theta}) is 
exactly invariant under the lead spin polarization $p$. In the antiparallel 
configuration, finite polarization introduces two incoherent tunneling 
channels per lead: a correct-spin channel with rate $p\,\Gamma_0$ and a 
wrong-spin channel with rate $(1-p)\,\Gamma_0$. The resulting master 
equation reads
\begin{align}
\frac{\partial\rho}{\partial\theta} &= 
-i\left[\frac{H_0}{\Omega},\rho\right] \notag \\
&+ \frac{p\,\Gamma_0}{\Omega}\,\mathcal{D}[d_{\delta+}]\rho
 + \frac{(1-p)\,\Gamma_0}{\Omega}\,\mathcal{D}[d_{\delta-}]\rho
\notag \\
&+ \frac{p\,\Gamma_0}{\Omega}\,\mathcal{D}[d_{\delta-}]\rho
 + \frac{(1-p)\,\Gamma_0}{\Omega}\,\mathcal{D}[d_{\delta+}]\rho,
\label{eq:lindblad_noisy}
\end{align}
where $\mathcal{D}[A]\rho = A\rho A^\dagger 
- \frac{1}{2}\{A^\dagger A,\rho\}$ denotes the Lindblad 
dissipator. Grouping the terms acting on the same 
spin channel:
\begin{align}
\frac{\partial\rho}{\partial\theta} &= 
-i\left[\frac{H_0}{\Omega},\rho\right] \notag \\
&+ \frac{\Gamma_0}{\Omega}
  \bigl[p + (1-p)\bigr]\,\mathcal{D}[d_{\delta+}]\rho \notag \\
&+ \frac{\Gamma_0}{\Omega}
  \bigl[(1-p) + p\bigr]\,\mathcal{D}[d_{\delta-}]\rho \notag \\
&= -i\left[\frac{H_0}{\Omega},\rho\right]
+ \frac{\Gamma_0}{\Omega}\,\mathcal{D}[d_{\delta+}]\rho
+ \frac{\Gamma_0}{\Omega}\,\mathcal{D}[d_{\delta-}]\rho,
\label{eq:lindblad_invariant}
\end{align}
which is identical to Eq.~(\ref{eq:lindblad_noisy}) 
with $p = 1$. The cancellation $p + (1-p) = 1$ is 
exact and holds for any value of $p \in [0,1]$: the 
spin-selectivity loss in one lead is exactly compensated 
by the wrong-spin leakage from the opposite lead, 
leaving the total dissipation rate per spin channel 
equal to $\Gamma_0$ regardless of $p$. This constitutes 
an analytical proof that the qubit dynamics is 
intrinsically insensitive to the degree of lead spin 
polarization in the antiparallel configuration, a result 
that is also confirmed numerically in Fig.~\ref{fig_fidelity_vs_polarization}.

It is worth noting that the $N = 10$ independent realizations employed 
here to estimate $\bar{F}$ and $\sigma_F$ have a direct experimental 
interpretation: each realization corresponds to a complete acquisition 
run of $R$ detection repetitions, and the standard deviation $\sigma_F$ 
quantifies the run-to-run variability that an experimentalist would 
observe upon repeating the full tomography protocol $N$ times. 
In this sense, $\sigma_F$ is not merely a computational artifact but 
a genuine prediction of the reproducibility of the protocol under 
realistic experimental conditions.

It is instructive to clarify what the GPR reconstruction can and cannot 
assess by itself. The GPR model has access to three types of information: 
(i)~the detected time points and their associated estimated density-matrix 
elements, which constitute the training data; (ii)~the smoothness structure 
of the underlying signal, encoded in the optimized kernel hyperparameters; 
and (iii)~its own uncertainty, expressed as $\pm 1\sigma$ bands that widen 
in time regions with sparse detection data. However, the GPR is 
intrinsically unaware of three physically important features: 
(i)~whether the reconstructed matrix is a valid density matrix, i.e., 
whether it is positive semidefinite with non-negative eigenvalues; 
(ii)~whether the reconstructed state is close to the ideal (decoherence-free) 
evolution; and (iii)~whether the input data are statistically reliable, 
as the GPR adjusts equally well to accurate and to noise-dominated data. 
This asymmetry motivates the introduction of the state fidelity 
$F(\rho, \sigma) = \bigl(\mathrm{Tr}\sqrt{\sqrt{\rho}\,\sigma\,\sqrt{\rho}}\bigr)^2$ 
as a complementary figure of merit. Unlike the $R^2_{\mathrm{GPR}}$ score, which measures 
how well the GPR fits the stochastic input data, the fidelity compares 
the reconstructed density matrix directly against the closed-system 
analytical solution $\sigma$, providing a global assessment of the 
tomography protocol that simultaneously accounts for the reconstruction 
quality and the decoherence-induced departure from the ideal state.

\begin{figure}[tb]
\centering
\includegraphics[width=0.70\linewidth]{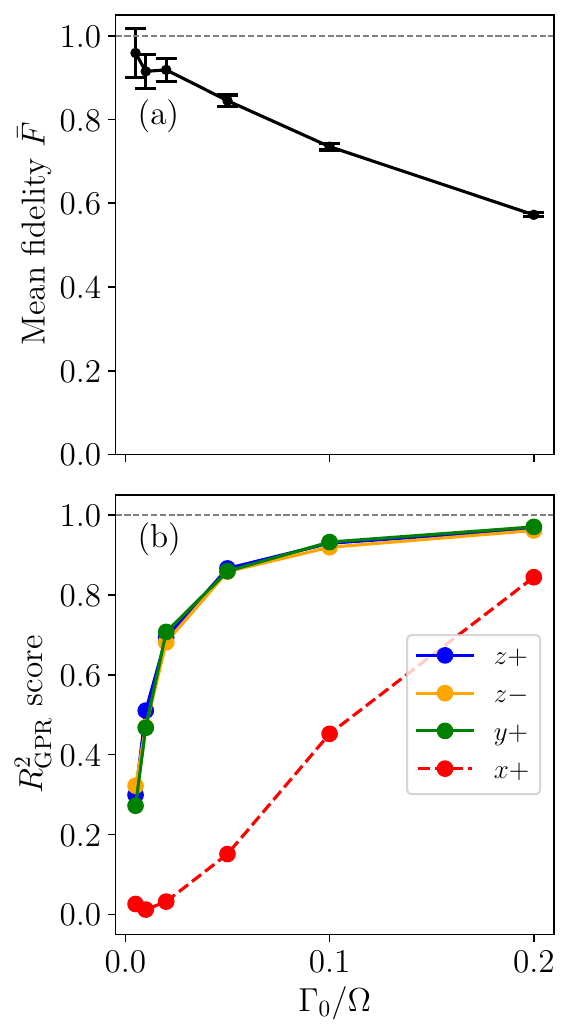}
\caption{(a) Mean state fidelity $\bar{F}$ and (b) GPR $R^2_{\mathrm{GPR}}$ score 
for each detection channel as a function of the tunneling rate $\Gamma_0/\Omega$, 
computed over $N = 10$ independent stochastic realizations with $R = 10{,}000$. 
In panel (b), the $R^2_{\mathrm{GPR}}$ score increases with $\Gamma_0$ for all 
channels, reflecting the improvement in GPR fit quality as more detection 
events become available. The channels $z+$ (blue), $z-$ (orange), and $y+$ 
(green) reach high $R^2_{\mathrm{GPR}}$ values already for moderate $\Gamma_0$, 
while the $x+$ channel (red dashed) starts significantly lower because its 
spin projection does not precess under the transverse field, resulting in a 
purely exponential decay with low temporal contrast. As $\Gamma_0$ increases, 
however, the exponential decay becomes more pronounced and better resolved by 
the stochastic detector, leading to a marked improvement in $R^2_{\mathrm{GPR}}$ 
for this channel as well. The two panels reveal an intrinsic trade-off: 
increasing $\Gamma_0$ improves the GPR fit quality (higher $R^2_{\mathrm{GPR}}$) 
by providing more detection events, but simultaneously increases decoherence 
and reduces the fidelity with respect to the closed-system reference. 
The value $\Gamma_0 = 0.05\,\Omega$ used in this work balances these two 
competing effects.
Parameters: $\Omega = 1~\mu\mathrm{eV}$ ($242~\mathrm{MHz}$), $R = 10{,}000$.\label{fig_tradeoff_gamma}}
\end{figure}

As shown in Fig.~\ref{fig_tradeoff_gamma}, these two metrics exhibit opposite 
trends with the tunneling rate $\Gamma_0$: while the $R^2_{\mathrm{GPR}}$ score increases 
with $\Gamma_0$ due to the larger number of detection events available for 
GPR training, the mean fidelity $\bar{F}$ decreases because stronger 
coupling to the leads induces greater decoherence, driving the open-system 
state further from the closed-system reference. A high $R^2_{\mathrm{GPR}}$ score therefore 
does not guarantee a high fidelity — the decoherence sets a physical upper 
bound on $\bar{F}$ that no machine-learning algorithm can overcome. 
The value $\Gamma_0 = 0.05\,\Omega$ used throughout this work represents 
a reasonable compromise between detection statistics and state quality.

\begin{figure}[tb]
\centering
\includegraphics[width=0.95\linewidth]{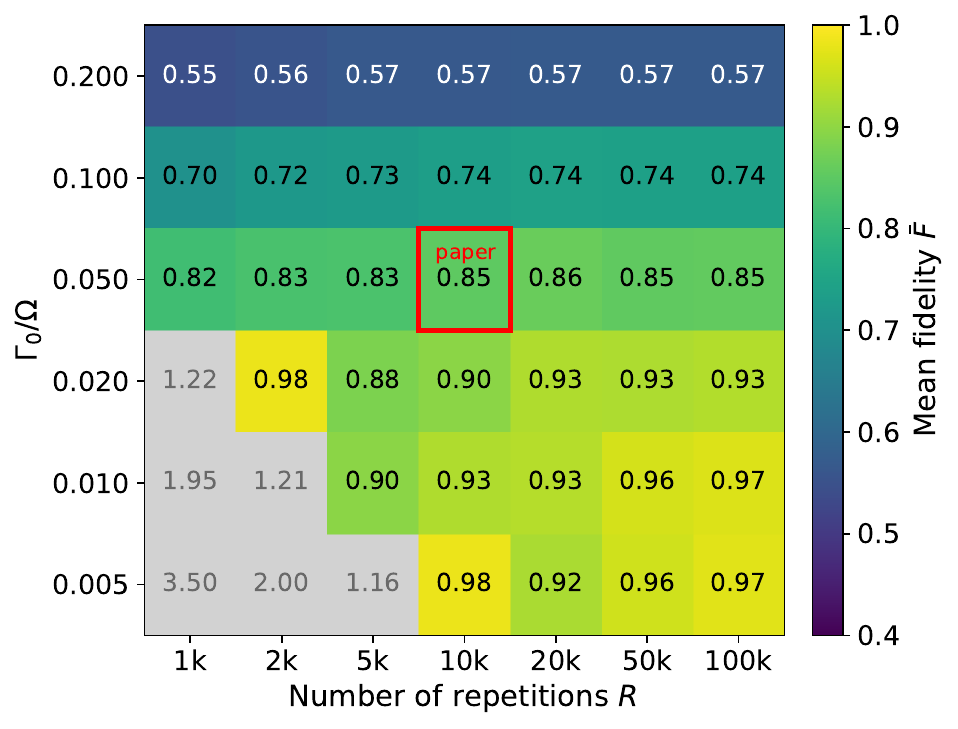}
\caption{Mean state fidelity $\bar{F}$ as a function of the tunneling 
rate $\Gamma_0/\Omega$ and the number of detection repetitions $R$, computed 
over $N = 10$ independent stochastic realizations per cell. The fidelity is 
computed with respect to the closed-system analytical solution. Gray cells 
indicate parameter regimes where the GPR reconstruction produces a 
non-physical density matrix (negative eigenvalues), reflecting insufficient 
detection data; the corresponding fidelity values fall outside $[0,1]$ 
and are excluded. The red box marks the parameter values used throughout 
this work ($\Gamma_0 = 0.05\,\Omega$, $R = 10{,}000$).
Parameters: $\Omega = 1~\mu\mathrm{eV}$ ($242~\mathrm{MHz}$), $N = 10$.\label{fig_heatmap_fidelity}}
\end{figure}

To provide a comprehensive picture of the protocol's validity across the 
full parameter space, Fig.~\ref{fig_heatmap_fidelity} shows the mean 
fidelity $\bar{F}$ as a function of both $\Gamma_0$ and $R$ simultaneously. 
The heatmap reveals a clear trade-off between decoherence (controlled by 
$\Gamma_0$) and detection statistics (controlled by $R$): high fidelity 
is achieved in the lower-right region (small $\Gamma_0$, large $R$), while 
low fidelity appears in the upper region (large $\Gamma_0$) regardless of $R$. 
Gray cells marked with a dash indicate parameter regimes where the GPR 
reconstruction produces a non-physical density matrix (with negative 
eigenvalues), corresponding to insufficient detection statistics. 
The fidelity values in these cells fall outside the physical range $[0,1]$ 
and are therefore excluded. The parameter values used in this work 
($\Gamma_0 = 0.05\,\Omega$, $R = 10{,}000$, highlighted in red) lie 
well within the valid reconstruction regime.

\section{Summary}
\label{sec:summary}

In this work, we have proposed and simulated a transport-based protocol for 
reconstructing the reduced density matrix of a single-electron spin qubit in 
a quantum dot coupled to ferromagnetic leads. By counting spin-polarized 
tunneling events along four distinct magnetization directions ($z+$, $z-$, 
$x+$, and $y+$), we estimate spin-resolved occupation probabilities that encode 
complete information about the qubit state, including both diagonal populations 
and off-diagonal coherences.

The detection statistics are generated via a stochastic simulation of the 
open-system dynamics governed by the Lindblad master equation, closely mimicking 
realistic experimental measurement outcomes. A Gaussian Process Regression (GPR) 
model is then employed as a post-processing tool to reconstruct smooth 
time-dependent density-matrix elements from the noisy counting statistics, 
providing not only a mean prediction but also quantified uncertainty bands 
that reflect the statistical quality of the detection data.

The robustness of the protocol was assessed along two axes. First, we showed 
analytically and confirmed numerically that the Lindblad master equation is 
exactly invariant under the lead spin polarization $p$: in the antiparallel 
configuration, the total dissipation rate per spin channel remains $\Gamma_0$ 
regardless of $p$, since the correct-spin contribution ($p\,\Gamma_0$) is 
exactly compensated by the wrong-spin leakage from the opposite lead 
($(1-p)\,\Gamma_0$). This constitutes an intrinsic robustness of the protocol 
against a key experimental imperfection in spintronic devices. Second, we 
demonstrated that the reconstruction fidelity $\bar{F}$ increases 
monotonically with the number of detection repetitions $R$, while the 
run-to-run standard deviation $\sigma_F$ decreases from $0.019$ at 
$R = 5{,}000$ to $0.005$ at $R = 50{,}000$, confirming that the protocol accuracy 
is directly controlled by the total acquisition time — an experimentally 
accessible parameter.

The protocol provides an alternative route to quantum state tomography that 
is naturally compatible with spintronic architectures based on electrically 
detected transport, without requiring optical access or resonant excitation. 
Our results demonstrate that combining stochastic spin-polarized detection 
with Gaussian Process Regression provides a powerful and statistically 
principled framework for quantum state reconstruction under experimentally 
feasible conditions. A systematic robustness analysis reveals an intrinsic 
trade-off between decoherence (controlled by the tunneling rate $\Gamma_0$) 
and detection statistics (controlled by the number of repetitions $R$): 
the GPR $R^2_{\mathrm{GPR}}$ score improves with $\Gamma_0$ due to the increasing number 
of detection events, while the mean fidelity $\bar{F}$ with respect to the 
closed-system reference decreases due to stronger decoherence. 
This trade-off, illustrated in Figs.~\ref{fig_tradeoff_gamma} 
and~\ref{fig_heatmap_fidelity}, defines the regime of validity of the 
protocol and provides practical guidance for choosing experimental parameters.

\begin{acknowledgments}
M.~B.~S. acknowledges financial support from CAPES. L.~S. and F.~M.~S. acknowledge financial support from the National Institute of Science and Technology for Applied Quantum Computing through Conselho Nacional de Desenvolvimento Científico e Tecnológico (CNPq), Grant No.~408884/2024-0.
\end{acknowledgments}

\appendix
\section{Lindblad Equation}
\label{appendix:openmeas}

To simulate the stochastic detection of spin-polarized electrons, we first need to generate the corresponding probability distributions for the electron spin within the quantum dot, taking into account the influence of all drain ferromagnetic leads. This requires modeling the quantum evolution of the system, which we begin by describing using the von Neumann equation ($\hbar = 1$):
\begin{equation}
	\dot{\rho}_{\mathrm{tot}}(t) = -i\left[H, \rho_{\mathrm{tot}}(t)\right],
\end{equation}
where $\rho_{\mathrm{tot}}(t)$ is the density matrix of the total system and $H$ is the full Hamiltonian given in Eq.~\eqref{H}.
To derive the Lindblad master equation, we move to the interaction picture, which simplifies the treatment of the system–environment coupling. In this representation, the density matrix transforms as
\begin{equation}
	\hat{\rho}_{\mathrm{tot}}(t) = e^{i h_f t} \rho_{\mathrm{tot}}(t) e^{-i h_f t},
\end{equation}
where $\hat{\rho}_{\mathrm{tot}}(t)$ is the density matrix in the interaction picture and $h_f$ is the free Hamiltonian of the system, given by:
	\[ h_f=H-V_{\delta}.\]
In the interaction picture, the time evolution of the total density matrix is governed by
\begin{equation}\label{rhohatdot1}
	\dot{\hat{\rho}}_{\mathrm{tot}}(t) = -i\left[\hat{V}_\delta (t), \hat{\rho}_{\mathrm{tot}}(t)\right],
\end{equation}
where $\hat{V}_\delta (t)$ is the coupling term in the interaction picture. Integrating Eq.~\eqref{rhohatdot1} and iterating up to second order in $\hat{V}_\delta(t)$ yields
\begin{eqnarray}\label{rhoQ_1}
	&& \dot{\hat{\rho}}_{\mathrm{tot}}(t) = -i [\hat{V}_\delta(t), \hat{\rho}_{\mathrm{tot}}(0)] \\
	&-& \int_0^t dt_1 \Big\{ \hat{V}_\delta(t) \hat{V}_\delta(t_1) \hat{\rho}_{\mathrm{tot}}(t_1) 
	- \hat{V}_\delta(t) \hat{\rho}_{\mathrm{tot}}(t_1) \hat{V}_\delta(t_1) \nonumber \\
	&&\quad - \hat{V}_\delta(t_1) \hat{\rho}_{\mathrm{tot}}(t_1) \hat{V}_\delta(t) 
	+ \hat{\rho}_{\mathrm{tot}}(t_1) \hat{V}_\delta(t_1) \hat{V}_\delta(t) \Big\}.\nonumber
\end{eqnarray}
To proceed, we invoke the Born–Markov approximation~\cite{BreuerPetruccione2002}, which assumes that the system and its environment remain approximately uncorrelated during the evolution (Born approximation) and that the reservoirs exhibit negligible memory effects (Markov approximation), leading to a time-local master equation. Accordingly, we express the total density matrix as
\begin{equation}
	\hat{\rho}_{\mathrm{tot}}(t) \approx \hat{\rho}(t) \otimes \rho_L \otimes \rho_R,
\end{equation}
where $\hat{\rho}(t)$ denotes the reduced density matrix of the quantum dot, and $\rho_{L(R)}$ are the stationary density matrices of the left (right) ferromagnetic leads.
To obtain the reduced dynamics of the quantum dot, we trace out the reservoir degrees of freedom, which include all modes of the $L$ and $R$ leads. The resulting equation for the reduced density matrix reads
\begin{eqnarray}\label{L}
	&& \hat{L}_\delta(t) = - \int_0^t dt_1  \sum_{\mathbf{k}} \sum_{\eta \in \{L, R\}} \sum_{\lambda \in \{+,-\}}   \Big\{  \\
	&& \phantom{+}  v_{\eta}(t) v_{\eta}^\star(t_1) \hat{d}_{\delta \lambda}(t) \hat{d}^\dagger_{\delta \lambda}(t_1) \hat{\rho}(t_1) \langle  \hat{c}_{\mathbf{k} \eta \delta \lambda}^\dagger(t) \hat{c}_{\mathbf{k} \eta \delta \lambda}(t_1) \rangle \nonumber \\
	&& + v_{\eta}^\star(t) v_{\eta}(t_1) \hat{d}_{\delta \lambda}^\dagger(t) \hat{d}_{\delta \lambda}(t_1) \hat{\rho}(t_1) \langle  \hat{c}_{\mathbf{k} \eta \delta \lambda}(t) \hat{c}_{\mathbf{k} \eta \delta \lambda}^\dagger(t_1) \rangle \nonumber \\
	&& - v_{\eta}(t) v_{\eta}^\star(t_1) \hat{d}_{\delta \lambda}(t) \hat{\rho}(t_1)  \hat{d}^\dagger_{\delta \lambda}(t_1) \langle  \hat{c}_{\mathbf{k} \eta \delta \lambda}(t_1) \hat{c}_{\mathbf{k} \eta \delta \lambda}^\dagger(t) \rangle \nonumber \\
	&& - v_{\eta}^\star(t) v_{\eta}(t_1) \hat{d}_{\delta \lambda}^\dagger(t) \hat{\rho}(t_1) \hat{d}_{\delta \lambda}(t_1)  \langle  \hat{c}_{\mathbf{k} \eta \delta \lambda}^\dagger(t_1) \hat{c}_{\mathbf{k} \eta \delta \lambda}(t)  \rangle \nonumber \\
	&& - v_{\eta}(t_1) v_{\eta}^\star(t) \hat{d}_{\delta \lambda}(t_1) \hat{\rho}(t_1)  \hat{d}^\dagger_{\delta \lambda}(t) \langle  \hat{c}_{\mathbf{k} \eta \delta \lambda}(t) \hat{c}_{\mathbf{k} \eta\delta  \lambda}^\dagger(t_1) \rangle \nonumber \\
	&& - v_{\eta}^\star(t_1) v_{\eta}(t) \hat{d}_{\delta \lambda}^\dagger(t_1) \hat{\rho}(t_1) \hat{d}_{\delta \lambda}(t)  \langle  \hat{c}_{\mathbf{k} \eta \delta \lambda}^\dagger(t) \hat{c}_{\mathbf{k} \eta \delta \lambda}(t_1)  \rangle \nonumber \\
	&& + v_{\eta}(t_1) v_{\eta}^\star(t) \hat{\rho}(t_1) \hat{d}_{\delta \lambda}(t_1)  \hat{d}^\dagger_{\delta \lambda}(t) \langle  \hat{c}_{\mathbf{k} \eta \delta \lambda}^\dagger(t_1) \hat{c}_{\mathbf{k} \eta \delta \lambda}(t) \rangle \nonumber \\
	&& + v_{\eta}^\star(t_1) v_{\eta}(t) \hat{\rho}(t_1) \hat{d}_{\delta \lambda}^\dagger(t_1) \hat{d}_{\delta \lambda}(t)  \langle  \hat{c}_{\mathbf{k} \eta \delta \lambda}(t_1) \hat{c}_{\mathbf{k} \eta \delta \lambda}^\dagger(t) \rangle \Big\}, \nonumber
\end{eqnarray}
where we have used the fact that the reservoir correlation functions satisfy$ \langle  \hat{c}_{\mathbf{k} \eta \delta \lambda}^\dagger(t) \hat{c}_{\mathbf{k'} \eta'  \delta \lambda'}(t_1) \rangle$ $\propto$ $\delta_{\mathbf{k}, \mathbf{k}'}$  $\delta_{ \eta,  \eta'}$  $\delta_{\lambda \lambda'}$ and similarly for the other expectation values. As the operator $\hat{c}_{\mathbf{k} \eta \delta \lambda}(t)$ evolves in the interaction picture simply as $\hat{c}_{\mathbf{k} \eta \delta \lambda}(t)=\hat{c}_{\mathbf{k} \eta \delta \lambda} e^{-i \varepsilon_{\mathbf{k} \eta \delta \lambda} t}$, Eq.~\eqref{L} can be cast into the form
\begin{eqnarray}\label{L1_2}
	&& \hat{L}_\delta(t) = - \int_0^t dt_1   \sum_{\mathbf{k}} \sum_{\eta \in \{L, R\}} \sum_{\lambda \in \{+,-\}}   \Big\{  \\
	&& \phantom{+}  v_{\eta}(t) v_{\eta}^\star(t_1) \hat{d}_{\delta \lambda}(t) \hat{d}^\dagger_{\delta \lambda}(t_1) \hat{\rho}(t_1)  n_{\mathbf{k} \eta \delta \lambda} e^{-i \varepsilon_{\mathbf{k} \eta \delta \lambda} (t_1-t)} \nonumber \\
	&& + v_{\eta}^\star(t) v_{\eta}(t_1) \hat{d}_{\delta \lambda}^\dagger(t) \hat{d}_{\delta \lambda}(t_1) \hat{\rho}(t_1) (1-n_{\mathbf{k} \eta \delta \lambda}) e^{-i \varepsilon_{\mathbf{k} \eta \delta \lambda} (t-t_1)} \nonumber \\
	&& - v_{\eta}(t) v_{\eta}^\star(t_1) \hat{d}_{\delta \lambda}(t) \hat{\rho}(t_1)  \hat{d}^\dagger_{\delta \lambda}(t_1) (1-n_{\mathbf{k} \eta \delta \lambda}) e^{-i \varepsilon_{\mathbf{k} \eta \delta \lambda} (t_1-t)}  \nonumber \\
	&& - v_{\eta}^\star(t) v_{\eta}(t_1) \hat{d}_{\delta \lambda}^\dagger(t) \hat{\rho}(t_1) \hat{d}_{\delta \lambda}(t_1)   n_{\mathbf{k} \eta \delta \lambda} e^{-i \varepsilon_{\mathbf{k} \eta \delta \lambda} (t-t_1)} \nonumber \\
	&& - v_{\eta}(t_1) v_{\eta}^\star(t) \hat{d}_{\delta \lambda}(t_1) \hat{\rho}(t_1)  \hat{d}^\dagger_{\delta \lambda}(t)  (1-n_{\mathbf{k} \eta \delta \lambda}) e^{-i \varepsilon_{\mathbf{k} \eta \delta \lambda} (t-t_1)} \nonumber \\
	&& - v_{\eta}^\star(t_1) v_{\eta}(t) \hat{d}_{\delta \lambda}^\dagger(t_1) \hat{\rho}(t_1) \hat{d}_{\delta \lambda}(t)  n_{\mathbf{k} \eta \delta \lambda} e^{-i \varepsilon_{\mathbf{k} \eta \delta \lambda} (t_1-t)} \nonumber \\
	&& + v_{\eta}(t_1) v_{\eta}^\star(t) \hat{\rho}(t_1) \hat{d}_{\delta \lambda}(t_1)  \hat{d}^\dagger_{\delta \lambda}(t)  n_{\mathbf{k} \eta \delta \lambda} e^{-i \varepsilon_{\mathbf{k} \eta \delta \lambda} (t-t_1)} \nonumber \\
	&& + v_{\eta}^\star(t_1) v_{\eta}(t) \hat{\rho}(t_1) \hat{d}_{\delta \lambda}^\dagger(t_1) \hat{d}_{\delta \lambda}(t)   (1-n_{\mathbf{k} \eta \delta \lambda}) e^{-i \varepsilon_{\mathbf{k} \eta \delta \lambda} (t_1-t)}  \Big\}, \nonumber
\end{eqnarray}
where $n_{\mathbf{k}\eta\delta\lambda}$ denotes the Fermi distribution function for the lead $\eta$ ($\eta \in \{L, R\}$). 

Now, we assume $n_{\mathbf{k}\eta\delta\lambda} = n_{\eta\delta\lambda} = 1$ for a source lead and $n_{\eta\delta\lambda} = 0$ for a drain lead. Integrating over energy using
\[
\sum_{\mathbf{k},\eta,\lambda} \longrightarrow \sum_{\eta,\lambda} \int d\varepsilon\, D_{\eta\delta\lambda},
\]
where $D_{\eta\delta\lambda}$ is the constant density of states for lead $\eta$ with spin projection $\lambda$ along axis $\delta$, we obtain
\begin{eqnarray}\label{L1_3}
	\hat{L}_\delta(t) &=& -\frac{1}{2} \sum_{\eta,\lambda} \Gamma_{\eta\delta\lambda}(t) \Big\{ \nonumber \\
	&& n_{\eta\delta\lambda} \big[ \hat{d}_{\delta\lambda}(t)\hat{d}^\dagger_{\delta\lambda}(t)\hat{\rho}(t) - \hat{d}^\dagger_{\delta\lambda}(t)\hat{\rho}(t)\hat{d}_{\delta\lambda}(t) \nonumber \\
	&& \phantom{xxx} - \hat{d}^\dagger_{\delta\lambda}(t)\hat{\rho}(t)\hat{d}_{\delta\lambda}(t) + \hat{\rho}(t)\hat{d}_{\delta\lambda}(t)\hat{d}^\dagger_{\delta\lambda}(t) \big] \nonumber \\
	&& + (1-n_{\eta\delta\lambda}) \big[ \hat{d}^\dagger_{\delta\lambda}(t)\hat{d}_{\delta\lambda}(t)\hat{\rho}(t) - \hat{d}_{\delta\lambda}(t)\hat{\rho}(t)\hat{d}^\dagger_{\delta\lambda}(t) \nonumber \\
	&& \phantom{xxxxx} - \hat{d}_{\delta\lambda}(t)\hat{\rho}(t)\hat{d}^\dagger_{\delta\lambda}(t) + \hat{\rho}(t)\hat{d}^\dagger_{\delta\lambda}(t)\hat{d}_{\delta\lambda}(t) \big] \Big\}, \nonumber
\end{eqnarray}
where
\[
\Gamma_{\eta\delta\lambda} = 2\pi |v_\eta|^2 D_{\eta\delta\lambda},
\]
is the tunneling rate between the quantum dot and lead $\eta$ for spin projection $\lambda$ along axis $\delta$.
In the configuration considered, both ferromagnetic leads are spin-polarized: the left lead transmits only spin-up electrons, while the right lead transmits only spin-down electrons, both along the same axis $\delta$. 
Furthermore, both leads act as drains, so $\Gamma_{L\delta +} = \Gamma_0$, $\quad \Gamma_{L \delta -} = 0$, $\quad \Gamma_{R \delta +} = 0$, $\quad \Gamma_{R \delta -} = \Gamma_0$,  and
$n_{L \delta +} = 0$, $\quad n_{R \delta -} = 0$. The general Lindblad master equation in the Schr\"odinger picture then reads:
\begin{eqnarray}\label{rho_lindblad}
	\dot{\rho}(t) &=& -i[H_0,\rho(t)] \nonumber \\
	&& -\frac{\Gamma_{L \delta +}}{2}\big[d_{\delta +}^\dagger d_{\delta +} \rho(t) - 2 d_{\delta +} \rho(t) d_{\delta +}^\dagger + \rho(t) d_{\delta +}^\dagger d_{\delta +} \big] \nonumber \\
	&& -\frac{\Gamma_{R \delta -}}{2}\big[d_{\delta -}^\dagger d_{\delta -} \rho(t) - 2 d_{\delta -} \rho(t) d_{\delta -}^\dagger + \rho(t) d_{\delta -}^\dagger d_{\delta -} \big], \nonumber
\end{eqnarray}
which can also be expressed in the dimensionless quantity $\theta = \Omega t$ as
\begin{eqnarray}
	&& \frac{\partial}{\partial \theta} \rho(\theta) = -i\left[\frac{H_0}{\Omega}, \rho(\theta)\right] \nonumber \\
	&& -\frac{1}{2} \frac{\Gamma_{L,\delta,+}}{\Omega} \big[ d_{\delta +}^\dagger d_{\delta +} \rho(\theta) - 2 d_{\delta +} \rho(\theta) d_{\delta +}^\dagger + \rho(\theta) d_{\delta +}^\dagger d_{\delta +} \big] \nonumber \\
	&& -\frac{1}{2} \frac{\Gamma_{R,\delta,-}}{\Omega} \big[ d_{\delta -}^\dagger d_{\delta -} \rho(\theta) - 2 d_{\delta -} \rho(\theta) d_{\delta -}^\dagger + \rho(\theta) d_{\delta -}^\dagger d_{\delta -} \big].\nonumber
\end{eqnarray}
This equation corresponds to the Eq.~\eqref{rho_lindblad_theta} from the main text.

\section{Detection Statistics and Regime of Validity}
\label{app:ndetec}

The reliability of the GPR reconstruction depends critically on the 
availability of sufficient detection data at each time point. 
We derive the mean number of detection events per time bin in four steps.

\textit{Step 1.} The probability of detecting a tunneling event in a 
real-time interval $\Delta t$ is
\begin{equation*}
p(t_i) = \frac{\Gamma_0}{\hbar}\,\Delta t,
\end{equation*}
where $\Gamma_0$ has units of energy ($\mu$eV) and $\hbar$ converts it 
to a rate.

\textit{Step 2.} We express $\Delta t$ in terms of the dimensionless 
time $\theta = \Omega t/\hbar$. Since $\Delta t = (\hbar/\Omega)\,\Delta\theta$,
substitution gives
\begin{equation*}
p(t_i) = \frac{\Gamma_0}{\hbar}\cdot\frac{\hbar}{\Omega}\,\Delta\theta 
= \frac{\Gamma_0}{\Omega}\,\Delta\theta \equiv p(\theta_i).
\end{equation*}
The $\hbar$ cancels exactly, leaving the dimensionless ratio 
$\Gamma_0/\Omega$ and a probability expressed entirely in terms of 
the dimensionless time $\theta_i$.

\textit{Step 3.} Over $R$ repetitions, the mean number of detections 
per bin is
\begin{equation}
\langle N_{\mathrm{det}} \rangle \approx R\,p(\theta_i) 
= R\,\frac{\Gamma_0}{\Omega}\,\Delta\theta 
= R\,\frac{\Gamma_0}{\Omega}\,\frac{2\pi}{N_T},
\label{eq:ndetec}
\end{equation}
where $\Delta\theta = 2\pi/N_T$ and $N_T = 100$ is the number of time bins. 
This is an upper-bound estimate since it assumes unit dot occupation 
$n(\theta_i) = 1$ at all times.

\textit{Step 4.} For the parameters used in this work 
($\Gamma_0 = 0.05\,\Omega$, $R = 10{,}000$, $N_T = 100$):
\begin{equation*}
\langle N_{\mathrm{det}} \rangle \approx 10{,}000 \times 0.05 \times 
\frac{2\pi}{100} = 10{,}000 \times 0.05 \times 0.0628 \approx 31,
\end{equation*}
confirming that the parameter regime used in this work lies well within 
the regime of reliable GPR reconstruction.

\begin{figure}[tb]
\centering
\includegraphics[width=0.95\linewidth]{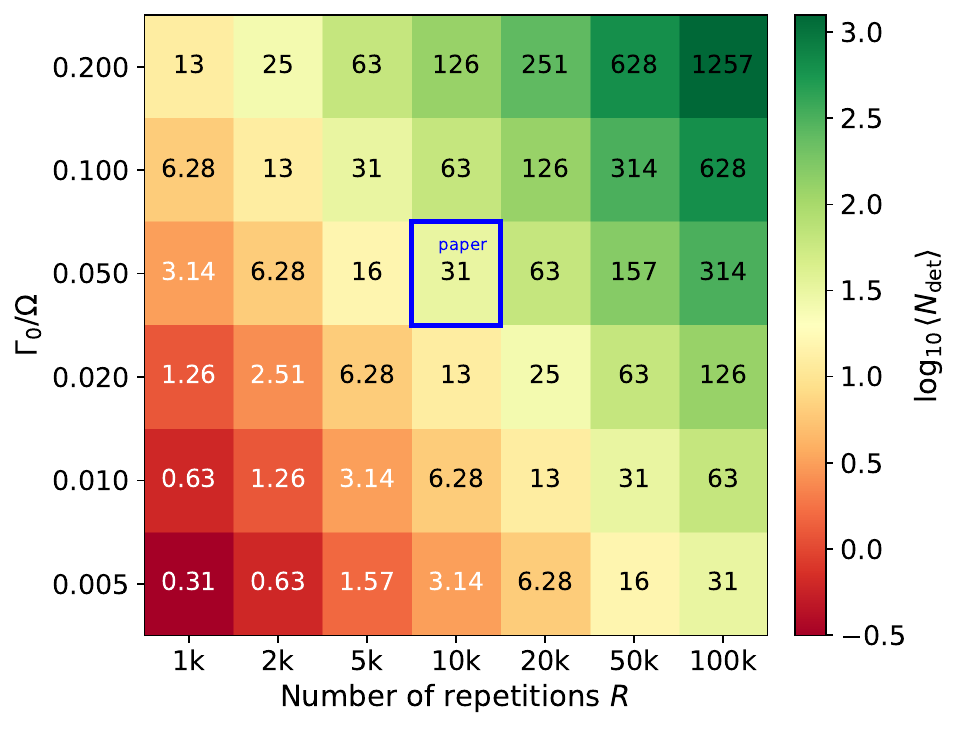}
\caption{Mean number of detection events per time bin, 
$\langle N_{\mathrm{det}}\rangle = R\,(\Gamma_0/\Omega)\,(2\pi/N_T)$
(upper-bound estimate assuming unit dot occupation $n(\theta_i) = 1$),
as a function of $\Gamma_0/\Omega$ and $R$. The color scale is 
logarithmic. The annotated values in each cell show the actual 
$\langle N_{\mathrm{det}}\rangle$. Cells with 
$\langle N_{\mathrm{det}}\rangle < 1$ correspond to the 
``insufficient data'' regime where the GPR reconstruction produces 
non-physical density matrices (gray cells in 
Fig.~\ref{fig_heatmap_fidelity}). The blue box marks the parameter 
values used throughout this work ($\Gamma_0 = 0.05\,\Omega$, 
$R = 10{,}000$, $\langle N_{\mathrm{det}}\rangle = 31$).
Parameters: $\Omega = 1~\mu\mathrm{eV}$ ($242~\mathrm{MHz}$), $N_T = 100$.\label{fig_ndetec_heatmap}}
\end{figure}

Fig.~\ref{fig_ndetec_heatmap} shows $\langle N_{\mathrm{det}} \rangle$ 
for the same parameter grid as the fidelity heatmap of 
Fig.~\ref{fig_heatmap_fidelity}. Cells with $\langle N_{\mathrm{det}}\rangle < 1$ 
(lower-left region, shown in red/orange) correspond to the gray 
``insufficient data'' cells in Fig.~\ref{fig_heatmap_fidelity}, 
confirming that the GPR reconstruction fails precisely when the 
detection statistics are too sparse. The parameter values used in 
this work ($\Gamma_0 = 0.05\,\Omega$, $R = 10{,}000$, blue box) 
yield $\langle N_{\mathrm{det}} \rangle = 31$, well above the 
threshold for reliable reconstruction.

\bibliography{Souza25}

\end{document}